\RequirePackage{fix-cm}
\documentclass[twocolumn,epjc3]{svjour3} 
\RequirePackage{cite}
\RequirePackage{epsfig}
\RequirePackage{amsmath} 
\RequirePackage{amssymb}
\RequirePackage{mathtools, cuted}
\smartqed  
\journalname{Eur. Phys. J. C}
\begin{document}

\title{Perturbatively stable observables in heavy-quark leptoproduction
}


\author{N.Ya.~Ivanov\thanksref{e1,addr1}
}

\thankstext{e1}{e-mail: nikiv@mail.yerphi.am}


\institute{Yerevan Physics Institute, Alikhanian Br.~2, 0036 Yerevan, Armenia \label{addr1}
}
\date{Received: date / Accepted: date}

\maketitle

\begin{abstract}
\noindent We study the perturbative and parametric stability of the QCD predictions 
for the Callan-Gross ratio $R(x,Q^2)=F_L/F_T$  and azimuthal $\cos(2\varphi)$ asymmetry, $A(x,Q^2)$, in heavy-quark leptoproduction.  We review the available theoretical  results for these quantities and conclude that, contrary to the production cross sections, the ratios $R(x,Q^2)$ and $A(x,Q^2)$ are stable under radiative QCD corrections in wide region of the variables $x$ and $Q^2$.  This implies that large radiative contributions to the structure functions cancel each other in the ratios $R(x,Q^2)$ and $A(x,Q^2)$ with good accuracy. 

\noindent Then we consider some experimental and phenomenological applications of the observed perturbative stability. We provide compact analytic predictions for $R(x,Q^2)$ and azimuthal $\cos(2\varphi)$ asymmetry in the case of low $x\ll 1$. It is demonstrated that our obtained results will be useful in the extraction of the structure functions from measurements of the reduced cross sections. Finally, we analyze the properties of $R(x,Q^2)$ and $A(x,Q^2)$ within the variable-flavor-number scheme (VFNS) of QCD. We conclude that the Callan-Gross ratio and azimuthal asymmetry are perturbatively stable but sensitive to resummation of the mass logarithms of the type $\alpha_{s}\ln\left( Q^{2}/m^{2}\right)$. For this reason, the quantities $R(x,Q^2)$ and $A(x,Q^2)$ will be good probes of the heavy-quark content of the proton. 
\end{abstract}

\section{Introduction}
\label{intro}
In principle, the solid theoretical justification of the QCD applicability to heavy-flavor production requires a detailed analysis of the convergence of the perturbative series for corresponding production cross sections. Presently, such analysis is below the horizon because the basic spin-averaged characteristics of heavy flavor photo- \cite{Ellis-Nason,Smith-Neerven},  electro- \cite{LRSN}, and hadro-production \cite{Nason-D-E-1,Nason-D-E-2,BKNS} are known exactly only up to the next-to-leading order (NLO) in $\alpha_s$.\footnote{Recently, some 25 years after the NLO results \cite{BKNS}, first complete next-to-next-to-leading order (NNLO) predictions for the heavy-quark pair hadroproduction were obtained \cite{Czakon-Mitov-1,Czakon-Mitov-2}.}  The problem is that these NLO corrections are large; they increase the leading-order (LO) predictions for both charm and bottom production cross sections by approximately a factor of two. Moreover, soft-gluon resummation of the threshold Sudakov logarithms indicates that higher-order contributions can also be substantial. (For details, see Refs.~\cite{Laenen-Moch,kid2}.) Perturbative instability leads to a high sensitivity of the theoretical calculations to standard uncertainties in the input QCD parameters. The total uncertainties associated with the unknown values of these parameters are so large that one can only estimate the order of magnitude of the perturbative QCD (pQCD) predictions for charm production cross sections in wide energy range \cite{Mangano-N-R,Frixione-M-N-R,R-Vogt,Moch2}.

Since the charm and bottom production cross sections are not perturbatively stable, it is of special interest to study those observables that are well-defined in pQCD. Measurements of such observables will provide, in particular, direct test of the conventional parton model based on pQCD. Moreover, as discussed below, some of the perturbatively stable quantities are sensitive to resummation of the mass logarithms and thus will be good probes of the heavy-quark densities in the proton. Experimental information about the heavy-quark content of the proton is necessary for construction of the appropriate variable-flavor-number  factorization scheme (VFNS) which may improve the convergence of the perturbative series \cite{ACOT,Collins}.

Nontrivial examples of the perturbatively stable observables were proposed in Refs.~\cite{we1,we2,we3,we4,we5,we6,we7,we8}, where the azimuthal $\cos(2\varphi)$ asymmetry, $A(x,Q^2)$, and Callan-Gross ratio, $R(x,Q^2)=F_L/F_T$, in heavy-quark leptoproduction were analyzed.\footnote{Well-known examples include the shapes of differential cross sections of heavy flavor production, which are sufficiently stable under radiative corrections. Note also the perturbative stability of the charge asymmetry in top-quark hadroproduction \cite{Almeida-S-V}.} 
In particular, radiative corrections to the azimuthal $\cos(2\varphi)$ asymmetry were considered in Refs.~\cite{we1,we2,we3,we4}. It was shown that, contrary to the production cross sections, the asymmetry is quantitatively well defined in pQCD: the contribution of the dominant photon-gluon fusion mechanism to $A(x,Q^2)$ is stable, both parametrically and perturbatively.

The perturbative and parametric stability of the ratio $R(x,Q^2)=F_L/F_T$ was discussed in Refs.~\cite{we7,we8}. It was shown that large perturbative contributions to the structure functions $F_T(x,Q^2)$ and $F_L(x,Q^2)$ cancel each other in their ratio $R(x,Q^2)$ with good accuracy. As a result, the NLO corrections to the LO photon-gluon fusion predictions for the Callan-Gross ratio are less than $10\%$ in a wide region of the variables $x$ and $Q^2$.

\begin{figure}
\begin{center}
\mbox{\epsfig{file=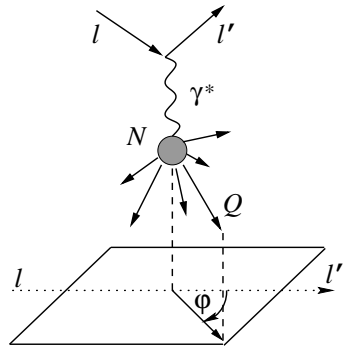,width=160pt}}
\caption{\label{Fg.1}\small Definition of the azimuthal angle $\varphi$ in the nucleon rest frame.}
\end{center}
\end{figure}

In the present paper, we continue the studies of perturbatively stable observables in heavy-quark leptoproduction,
\begin{equation}
\ell(l )+N(p)\rightarrow \ell(l -q)+Q(p_{Q})+X[\bar{Q}](p_{X}). \label{1}
\end{equation}
Neglecting the contribution of $Z$-boson exchange, the azimuth-dependent cross section of the reaction (\ref{1}) can be written as
\begin{eqnarray}
\frac{\mathrm{d}^{3}\sigma_{lN}}{\mathrm{d}x\mathrm{d}Q^{2}\mathrm{d}\varphi }&=&\frac{2\alpha^{2}_{em}}{Q^4}
\frac{y^2}{1-\varepsilon}\Bigl[ F_{T}( x,Q^{2})+ \varepsilon F_{L}(x,Q^{2}) \Bigr. \nonumber \\
&&+\Bigl. \varepsilon F_{A}( x,Q^{2})\cos 2\varphi  \nonumber \\
&&+2\sqrt{\varepsilon(1+\varepsilon)} F_{I}( x,Q^{2})\cos \varphi\Bigr], \label{2}
\end{eqnarray}
where $\alpha_{\mathrm{em}}$ is Sommerfeld's fine-structure constant, \\
$F_{2}(x,Q^2)=2x(F_{T}+F_{L})$, the quantity $\varepsilon$ measures the degree of the longitudinal polarization of the virtual photon in the Breit frame \cite{dombey}, $\varepsilon=\frac{2(1-y)}{1+(1-y)^2}$, and the kinematic variables are defined by
\begin{eqnarray}
\bar{S}=2\left( \ell\cdot p\right),\, \qquad &Q^{2}=-q^{2},\qquad \quad &x=\frac{Q^{2}}
{2p\cdot q},  \nonumber \\
y=\frac{p\cdot q}{p\cdot \ell },\qquad \quad ~ &Q^{2}=xy\bar{S},\qquad \quad &\xi=
\frac{Q^2}{m^2}.  \label{3}
\end{eqnarray}
In the nucleon rest frame, the azimuth $\varphi$ is the angle between the lepton scattering plane and the heavy quark production plane, defined by the exchanged photon and the detected quark $Q$ (see Fig.~\ref{Fg.1}). The covariant definition of $\varphi $ is
\begin{eqnarray}
\cos \varphi &=&\frac{r\cdot n}{\sqrt{-r^{2}}\sqrt{-n^{2}}},\, 
\sin \varphi =\frac{Q^{2}\sqrt{1/x^{2}+4m_{N}^{2}/Q^{2}}}{2\sqrt{-r^{2}}%
\sqrt{-n^{2}}}n\cdot \ell , \nonumber \\  
r^{\mu } &=&\varepsilon ^{\mu \nu \alpha \beta }p_{\nu }q_{\alpha }\ell _{\beta },\quad \,
 n^{\mu }=\varepsilon ^{\mu \nu \alpha \beta }q_{\nu }p_{\alpha }p_{Q\beta }.  \label{5}
\end{eqnarray}
In Eqs.~(\ref{3}) and (\ref{5}), $m$ and $m_{N}$ are the masses of the heavy quark and the target, respectively.

The Callan-Gross ratio, $R(x,Q^{2})$, and azimuthal $\cos(2\varphi)$ asymmetry, $A(x,Q^{2})$, are defined as
\begin{equation}\label{6}
R(x,Q^{2})=\frac{F_{L}}{F_{T}}(x,Q^{2}),\,\,\,\,\,  A(x,Q^{2})=2x\frac{F_{A}}{F_{2}}(x,Q^{2}).
\end{equation}

In this paper, we first review the available theoretical results for the quantities $R(x,Q^{2})$ and $A(x,Q^{2})$ adding for completeness the ingredients missed in previous analyses. In  particular, in Refs.~\cite{we7,we8}, only the contributions of the photon-gluon fusion mechanism to $R(x,Q^2)$ were considered at both LO and NLO. Now, using the explicit NLO results \cite{LRSN,Blumlein}, we provide the complete NLO predictions which include the contributions of both the photon-gluon, $\gamma ^{*}g\to Q\bar{Q}(g)$, and photon-(anti)quark, $\gamma ^{*}q\to Q\bar{Q}q$, fusion components. The complete ${\cal O}(\alpha_{s}^2)$ corrections to $R(x,Q^2)$ do not exceed 10--15$\%$ in the energy range $x>10^{-4}$.

Presently, the exact NLO predictions for the azimuth dependent structure function $F_{A}(x,Q^{2})$ are not available. For this reason, we use the soft-gluon approximation to estimate the radiative corrections to $F_{A}(x,Q^{2})$. 
Our analysis shows that the NLO soft-gluon predictions for $A(x,Q^{2})$ affect the LO results by less than a few percent at $Q^2 \lesssim m^2$ and $x\gtrsim 10^{-2}$. 

Note also that both the LO and NLO predictions for the Callan-Gross ratio and azimuthal asymmetry are sufficiently insensitive, to within ten percent, to standard uncertainties in the QCD input parameters $\mu_{F}$, $\mu_{R}$, $\Lambda_{\mathrm{QCD}}$, and the parton distribution functions (PDFs). 

Then we consider some experimental and phenomenological applications of the observed perturbative stability. We derive the compact analytic formulae for the hadron-level azimuthal asymmetry and Callan-Gross ratio in the limit of low $x\ll 1$. It is shown that our analytic LO results for $A(x\to 0,Q^{2})$ and $R(x\to 0,Q^2)$ are stable not only under the NLO corrections to the partonic cross sections, but also under the DGLAP  \cite{DGLAP1,DGLAP2,DGLAP3} evolution of the gluon PDF.

As to the experimental applications, our compact LO formula for $R(x\to 0,Q^2)$ conveniently reproduce the last HERA results for $F_2^c(x,Q^2)$ and $F_2^b(x,Q^2)$ obtained by H1  Collaboration \cite{H1HERA1} with the help of more cumbersome NLO estimations of $F_{L}(x,Q^2)$. Analytic predictions for $A(x\to 0,Q^{2})$ will be useful in extraction of the azimuthal asymmetries from the incoming COMPASS results as well as from future data on heavy-quark leptoproduction at the proposed EIC \cite{EIC} and LHeC \cite{LHeC,LHeC2} colliders at BNL/JLab and CERN, correspondingly.

Finally, we analyze the properties of $R(x,Q^2)$ and $A(x,Q^2)$ within the variable- flavor- number scheme (VFNS) of QCD. These quantities seems to be very promising probes of the heavy-quark densities in the proton. This is because the Callan-Gross ratio and azimuthal asymmetry are perturbatively stable but sensitive to resummation of the mass logarithms of the type $\alpha_{s}\ln\left( Q^{2}/m^{2}\right)$. Our analysis shows that resummation of the mass logarithms leads to reduction of the ${\cal O}(\alpha_{s})$ predictions for $A(x,Q^2)$ and $R(x,Q^2)$ by $(30$--$50)\%$ at $x\sim 10^{-2}$--$10^{-1}$ and $Q^2\gg m^2$.\footnote{At ${\cal O}(\alpha_{s}^2)$, the corresponding reduction of the finite-flavor-number scheme predictions for $R(x,Q^2)$ is estimated to be about 20$\%$.} We conclude that the ratios $R(x,Q^2)$ and $A(x,Q^2)$ will be good probes of the heavy-quark content of the proton.

This paper is organized as follows. In Section~\ref{NLO}, we analyze the exact NLO results
for the Callan-Gross ratio. The soft-gluon contributions to $A(x,Q^{2})$ are investigated in Section~\ref{SGR}. The analytic LO results for the ratios $R(x,Q^{2})$ and $A(x,Q^{2})$ at low $x$ are discussed in Section~\ref{analytic}. In Secton~\ref{resum}, we consider the resummation of the mass logarithms of the type $\alpha_{s}\ln\left( Q^{2}/m^{2}\right)$ for the Callan-Gross ratio and azimuthal $\cos(2\varphi)$ asymmetry. 

\section{\label{NLO} Exact NLO predictions for the Callan-Gross ratio $R(x,Q^2)$}

\begin{figure*}
\begin{center}
\mbox{\epsfig{file=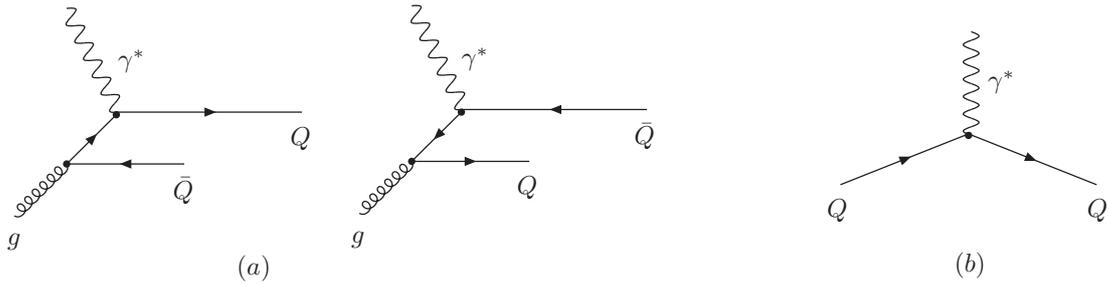,width=450pt}}
\end{center}
\caption{\label{Fg.2}\small LO Feynman diagrams of the photon-gluon fusion (a) and photon-quark scattering (b).}
\end{figure*}

At leading order, ${\cal O}(\alpha_{s})$, leptoproduction of heavy 
flavors proceeds through the photon-gluon fusion (GF) mechanism,
\begin{equation} \label{7}
\gamma ^{*}(q)+g(k_{g})\rightarrow Q(p_{Q})+\bar{Q}(p_{\bar{Q}}).
\end{equation}
The relevant Feynman diagrams are depicted in Fig.~\ref{Fg.2}\emph{a}.
The corresponding $\gamma ^{*}g$ cross sections, $\hat{\sigma}_{k,g}^{(0)}(z,\lambda)$ ($k=2,L,A,I$), have the form \cite{LW1}:
\begin{eqnarray}
\hat{\sigma}_{2,g}^{(0)}(z,\lambda)&=&\frac{\alpha_{s}}{2\pi}\hat{\sigma}_{B}(z)
\Bigl\{\bigl[(1-z)^{2}+z^{2}+4\lambda z(1-3z) \nonumber\\
&&-8\lambda^{2}z^{2}\bigr]
\ln\frac{1+\beta_{z}}{1-\beta_{z}}  \nonumber\\
&&-\left[1+4z(1-z)(\lambda-2)\right]\beta_{z}\Bigr\}, \label{8} \\ 
\hat{\sigma}_{L,g}^{(0)}(z,\lambda)&=&\frac{2\alpha_{s}}{\pi}\hat{\sigma}_{B}(z)z
\Bigl\{-2\lambda z\ln\frac{1+\beta_{z}}{1-\beta_{z}}+\left(1-z\right)\beta_{z}\Bigr\},\nonumber\\ 
\hat{\sigma}_{A,g}^{(0)}(z,\lambda)&=&\frac{\alpha_{s}}{\pi}\hat{\sigma}_{B}(z)z
\Bigl\{2\lambda\left[1-2z(1+\lambda)\right]\ln\frac{1+\beta_{z}}{1-\beta_{z}} \nonumber  \\
&&+(1-2\lambda)(1-z)\beta_{z}\Bigr\}, \nonumber  \\
\hat{\sigma}_{I,g}^{(0)}(z,\lambda)&=&0,  \nonumber
\end{eqnarray}
with $\hat{\sigma}_{B}(z)=(2\pi)^2e_{Q}^{2}\alpha_{\mathrm{em}}\,z/Q^{2}$, 
where $e_{Q}$ is the electric charge of quark $Q$ in units of the positron charge and  
$\alpha_{s}\equiv\alpha_{s}(\mu_R^2)$ is the strong-coupling constant. 
In Eqs.~(\ref{8}), we use the following definition of partonic kinematic variables:
\begin{equation}\label{9}
z=\frac{Q^{2}}{2q\cdot k_{g}},\qquad\lambda =\frac{m^{2}}{Q^{2}}, \qquad
\beta_{z}=\sqrt{1-\frac{4\lambda z}{1-z}}.
\end{equation}
The hadron-level cross sections, $\sigma_{k,GF}(x,Q^2)$ ($k=2,L,A,I$), corresponding to the GF subprocess, have the form
\begin{equation}\label{10}
\sigma_{k,GF}(x,Q^2)=\int_{x(1+4\lambda)}^{1}\mathrm{d}z\,g(z,\mu_{F})
\hat{\sigma}_{k,g}\left(x/z,\lambda,\mu_{F}\right),
\end{equation}
where $g(z,\mu_{F})$ is the gluon PDF of the proton. 

The leptoproduction cross sections $\sigma_{k}(x,Q^2)$ are related to the structure functions $F_{k}(x,Q^2)$ as follows:
\begin{eqnarray}
F_{k}(x,Q^2) &=&\frac{Q^{2}}{8\pi^{2}\alpha_{\mathrm{em}}x}\sigma_{k}(x,Q^2)
\qquad (k=T,L,A,I),  \nonumber \\
F_{2}(x,Q^2) &=&\frac{Q^{2}}{4\pi^{2}\alpha_{\mathrm{em}}}\sigma_{2}(x,Q^2), \label{11}
\end{eqnarray}
where $\sigma_{2}(x,Q^2)=\sigma_{T}(x,Q^2)+\sigma_{L}(x,Q^2)$.

At NLO, ${\cal O}(\alpha_{s}^2)$, the contributions of both the photon-gluon, $\gamma ^{*}g\to Q\bar{Q}(g)$, and photon-(anti)quark, $\gamma ^{*}q\to Q\bar{Q}q$, fusion components are usually presented in terms of the dimensionless coefficient functions $c_{k}^{(n,l)}(z,\lambda)$ as
\begin{eqnarray}
\hat{\sigma}_{k}(z,\lambda,\mu^{2})&=&\frac{e_{Q}^{2}\alpha_{\mathrm{em}}\alpha_{s}}{m^{2}}
\Bigl\{ c_{k}^{(0,0)}(z,\lambda)+4\pi\alpha_{s}\Bigl[c_{k}^{(1,0)}(z,\lambda)  \nonumber \\
&&+c_{k}^{(1,1)}(z,\lambda)\ln\frac{\mu^{2}}{m^{2}}
\Bigr]\Bigr\}+{\cal O}(\alpha_{s}^3), \label{12}
\end{eqnarray}
where we identify $\mu=\mu_{F}=\mu_{R}$.

The coefficients $c_{k,g}^{(1,1)}(z,\lambda)$ and $c_{k,q}^{(1,1)}(z,\lambda)$ ($k=T,L,\\A,I$) of the $\mu$-dependent logarithms can be evaluated explicitly using renormalization group arguments \cite{Ellis-Nason,LRSN}. The results of direct calculations of the coefficient functions $c_{k,g}^{(1,0)}(z,\lambda)$ and $c_{k,q}^{(1,0)}(z,\lambda)$ for $k=T,L$ are presented in Refs.~ \cite{LRSN,Blumlein}. Using these NLO predictions, we analyze the $Q^2$ dependence of the ratio $R(x,Q^2)=F_L/F_T$ at fixed values of $x$. 

\begin{figure*}
\begin{center}
\begin{tabular}{cc}
\mbox{\epsfig{file=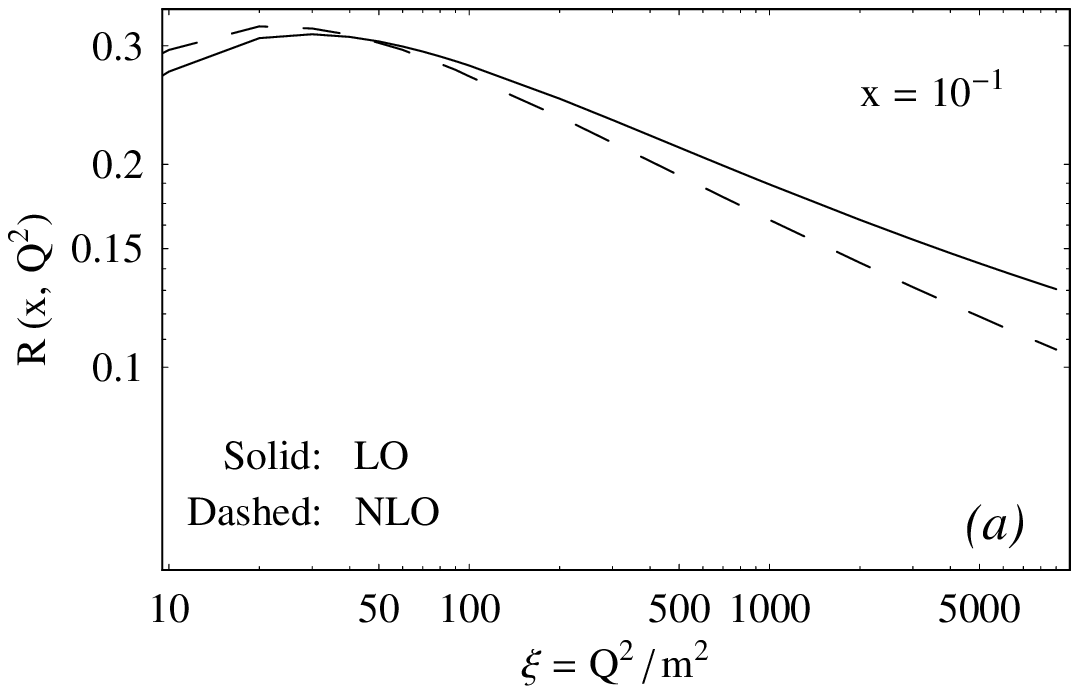,width=230pt}}
& \mbox{\epsfig{file=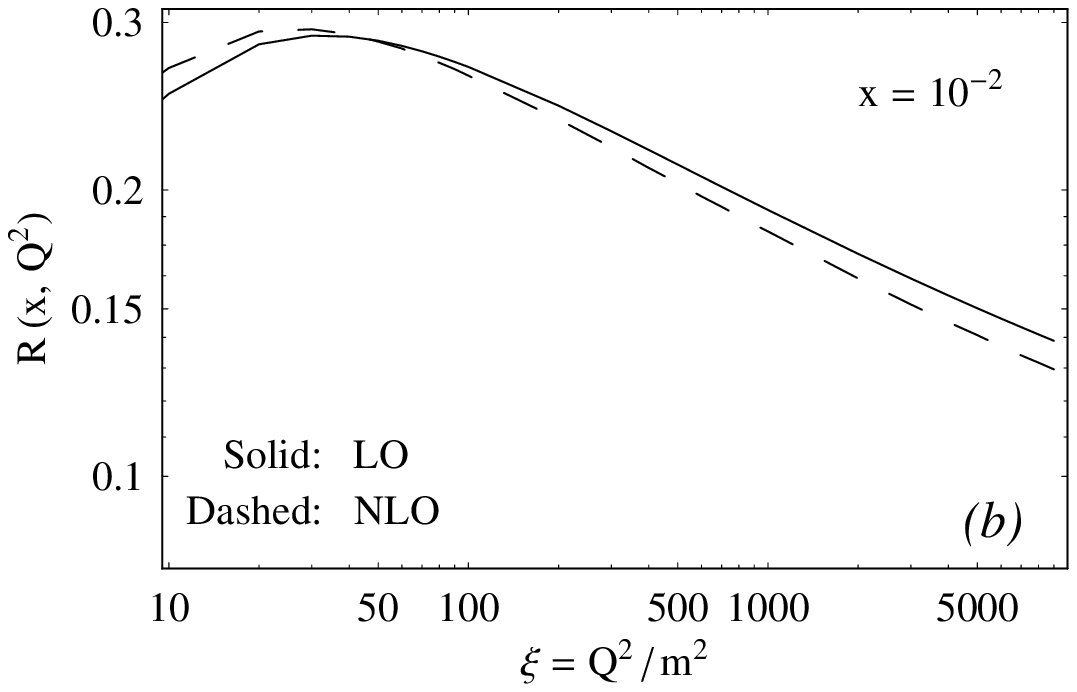,width=230pt}}\\
\mbox{\epsfig{file=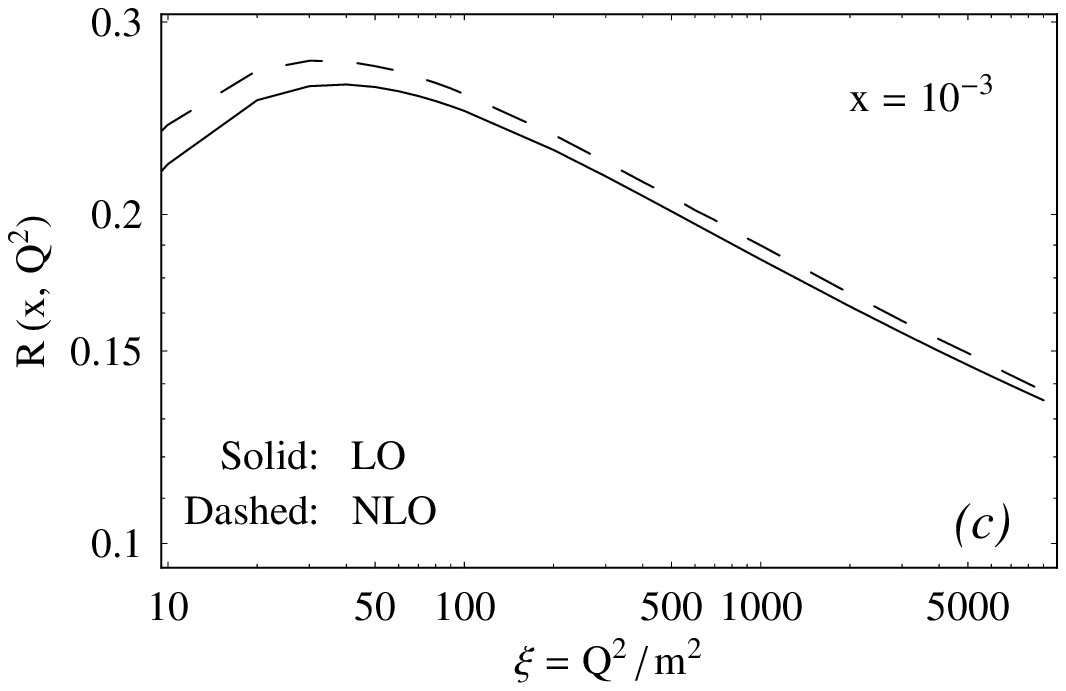,width=230pt}}
& \mbox{\epsfig{file=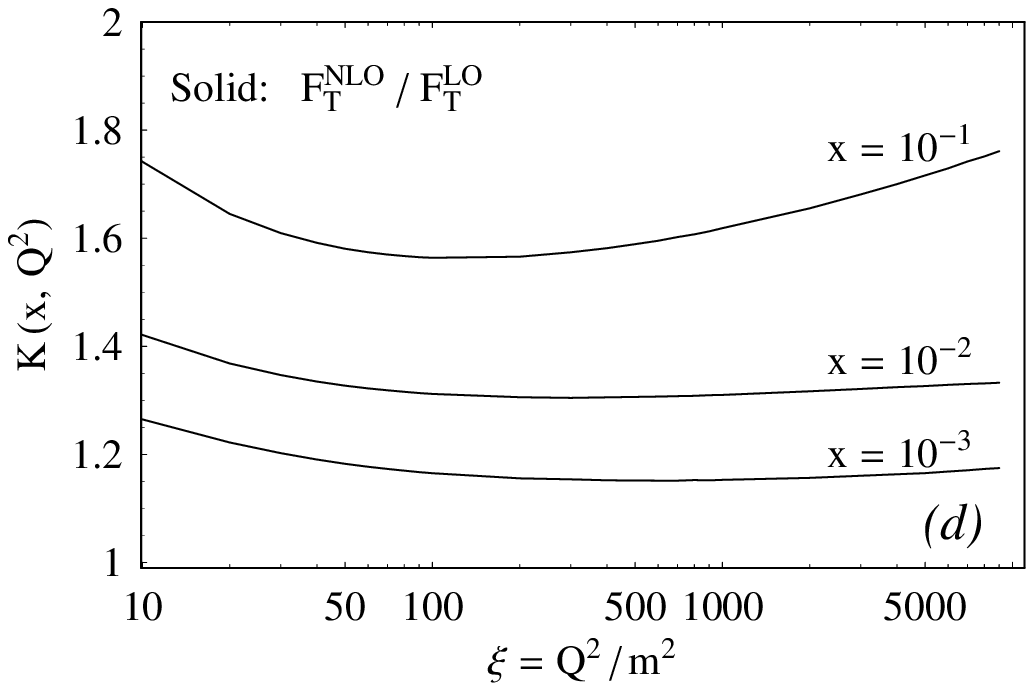,width=230pt}}\\
\end{tabular}
\caption{\label{Fg.3}\small $(a)$, $(b)$ and $(c)$ \emph{panels:} $Q^2$ dependence of the LO (solid curves) and NLO (dashed curves) predictions for the Callan-Gross ratio, $R(x,Q^2)=F_L/F_T$, in charm leptoproduction at $x=10^{-1}$, $10^{-2}$ and $10^{-3}$. \emph{$(d)$~panel:} $Q^2$ dependence of the $K$ factor for the transverse structure function,
$K(x,Q^2)=F_T^{\mathrm{NLO}}/F_T^{\mathrm{LO}}$, at the same values of $x$.}
\end{center}
\end{figure*}

The panels $(a)$, $(b)$ and $(c)$ of Fig.~\ref{Fg.3} show the NLO predictions for Callan-Gross ratio $R(x,Q^2)$ in charm leptoproduction as a function of $\xi=Q^2 /m^2$ at $x=10^{-1}$, $10^{-2}$ and $10^{-3}$, correspondingly. In our calculations, we use the CTEQ6M parametrization of the PDFs together with the values $m_c=1.3$~GeV and $\Lambda=326$~MeV \cite{CTEQ6}.\footnote{Note that we convolute the NLO CTEQ6M distribution functions with both the LO and NLO partonic cross sections that makes it possible to estimate directly the degree of stability of the pQCD predictions under radiative corrections.} Unless otherwise stated, we use $\mu=\sqrt{4m_c^{2}+Q^{2}}$ throughout this paper.

For comparison, the panel $(d)$ of Fig.~\ref{Fg.3} shows the $Q^2$ dependence of the QCD
correction factor for the transverse structure function,
$K(x,Q^2)=F_T^{\mathrm{NLO}}/F_T^{\mathrm{LO}}$. One can see that sizable radiative corrections to the structure functions $F_T(x,Q^2)$ and $F_L(x,Q^2)$ cancel each other in their ratio $R(x,Q^2)=F_L/F_T$ with good accuracy. As a result, the NLO contributions to the ratio $R(x,Q^2)$ are of the order of $10\%$ for $x > 10^{-4}$.

Another remarkable property of the Callan-Gross ratio closely related to fast
perturbative convergence is its parametric stability.\footnote{Of course, parametric
stability of the fixed-order results does not imply a fast convergence of the
corresponding series. However, a fast convergent series must be parametrically stable. In
particular, it must exhibit feeble $\mu_{F}$ and $\mu_{R}$ dependences.}
Our analysis shows that the fixed-order predictions for the ratio $R(x,Q^2)$ are less sensitive to standard uncertainties in the QCD input parameters than the corresponding ones for the production cross sections. For instance, sufficiently above the production threshold, changes of
$\mu$ in the range $(1/2)\sqrt{4m_{c}^{2}+Q^{2}}<\mu <2 \sqrt{4m_{c}^{2}+Q^{2}}$
only lead to $10\%$ variations of $R(x,Q^{2})$ at NLO. For comparison, at
$x=0.1$ and $\xi = 4.4$, such changes of $\mu$ affect the NLO predictions for the
quantities $F_{T}(x,Q^2)$ and $R(x,Q^{2})$ in charm leptoproduction by more than $100\%$
and less than $10\%$, respectively.

Keeping the value of the variable $Q^{2}$ fixed, we analyze the dependence of the pQCD
predictions on the uncertainties in the heavy-quark mass. We observe that changes of the charm-quark mass in the interval 1.3 $<m_{c}<1.7$ GeV affect the Callan-Gross ratio by (2--3)\% at $Q^{2}=10$ GeV$^2$ and $x<10^{-1}$. The corresponding variations of the structure functions $F_T(x,Q^2)$ and $F_L(x,Q^2)$ are about 20\%.
We also verified that the recent CTEQ versions \cite{CTEQ6,CT10,CT14} \footnote{For a review of the present status of all currently available PDF sets, see Ref.~\cite{PDF-LHC}.} of the PDFs lead to NLO predictions for $R(x,Q^{2})$ that coincide with each other with an accuracy of about
$5\%$ at $10^{-3}\leq x< 10^{-1}$.

\section{\label{SGR} Soft-gluon corrections to the azimuthal asymmetry $A(x,Q^{2})$ at NLO}

Presently, the exact NLO predictions for the azimuth dependent structure function $F_{A}(x,Q^{2})$ are not available. For this reason, we consider the NLO predictions for the azimuthal $\cos(2\varphi)$ asymmetry within the soft-gluon approximation. For the reader's convenience, we collect the final results for the parton-level GF cross sections to the next-to-leading logarithmic (NLL) accuracy. More details may be found in
Refs.~\cite{Laenen-Moch,we2,we4,we7}.

At NLO, photon-gluon fusion receives contributions from the virtual ${\cal
O}(\alpha_{\mathrm{em}}\alpha_{s}^{2})$ corrections to the Born process~(\ref{7}) and
from real-gluon emission,
\begin{equation}  \label{13}
\gamma ^{*}(q)+g(k_{g})\rightarrow Q(p_{Q})+\bar{Q}(p_{\bar{Q}})+g(p_{g}).
\end{equation}
The partonic invariants describing the single-particle inclusive (1PI) kinematics are
\begin{eqnarray}
s^{\prime }&=&2q\cdot k_{g}=\zeta S^{\prime }, \quad  \, t_{1}=\left(
k_{g}-p_{Q}\right) ^{2}-m^{2}=\zeta T_{1},  \nonumber \\
s_{4}&=&s^{\prime }+t_{1}+u_{1},\quad  \, ~ u_{1}=\left(
q-p_{Q}\right) ^{2}-m^{2}=U_{1}, \label{14}
\end{eqnarray}
where $\zeta$ is defined through $\vec{k}_{g}= \zeta\vec{p}$, $s^{\prime}=s+Q^{2}$, and $s_{4}$ measures the inelasticity of the reaction (\ref{13}). The corresponding 1PI hadron-level variables describing the reaction (\ref{1}) are
\begin{eqnarray}
S^{\prime }&=&2q\cdot p=S+Q^{2},\qquad  T_{1}=\left( p-p_{Q}\right)
^{2}-m^{2},  \nonumber \\
S_{4}&=&S^{\prime }+T_{1}+U_{1},\qquad \quad U_{1}=\left( q-p_{Q}\right)
^{2}-m^{2}. \label{15}
\end{eqnarray}

The exact NLO calculations of unpolarized heavy-quark production \cite{Ellis-Nason,Smith-Neerven,LRSN,Nason-D-E-1} show that, near the partonic
threshold, a strong logarithmic enhancement of the cross sections takes place in the
collinear, $|\vec{p}_{g,T}|\to 0$, and soft, $|\vec{p}_{g}|\to 0$,
limits. This threshold (or soft-gluon) enhancement is of universal nature in perturbation
theory and originates from an incomplete cancellation of the soft and collinear
singularities between the loop and the bremsstrahlung contributions. Large leading and
next-to-leading threshold logarithms can be resummed to all orders of the perturbative
expansion using the appropriate evolution equations
\cite{Contopanagos-L-S}. The analytic results for the resummed
cross sections are ill-defined due to the Landau pole in the coupling constant
$\alpha_{s}$. However, if one considers the obtained expressions as generating functionals
and re-expands them at fixed order in $\alpha_{s}$, no
divergences associated with the Landau pole are encountered.

Soft-gluon resummation for the photon-gluon fusion was performed in
Ref.~\cite{Laenen-Moch} and confirmed in Refs.~\cite{we2,we4}. To NLL accuracy, the
perturbative expansion for the partonic cross sections,
$\mathrm{d}^{2}\hat{\sigma}_{k}(s^{\prime},t_{1},u_{1})/(\mathrm{d}t_{1}\,
\mathrm{d}u_{1})$ ($k=T,L,A,I$), can be written in factorized form as
\begin{eqnarray} 
s^{\prime 2}\frac{\mathrm{d}^{2}\hat{\sigma}_{k}}{\mathrm{d}t_{1}\mathrm{d}u_{1}}( s^{\prime
},&&\!\! t_{1},u_{1})=B_{k}^{\mathrm{ Born}}( s^{\prime },t_{1},u_{1})\Biggl[\delta
(s^{\prime }+t_{1}+u_{1}) \nonumber \\
&&+\sum_{n=1}^{\infty } \left( \frac{\alpha
_{s}C_{A}}{\pi}\right)^{n}K^{(n)}( s^{\prime },t_{1},u_{1})\Biggr]. \label{16}
\end{eqnarray}

The functions $K^{(n)}( s^{\prime },t_{1},u_{1}) $ in Eq.~(\ref{16}) originate from the collinear and soft limits. Since the azimuthal angle $\varphi $ is the same for both $\gamma ^{*}g$ and $Q\bar{Q}$ center-of-mass systems in these limits, the functions 
$K^{(n)}( s^{\prime },t_{1},u_{1}) $ are also the same for all $\hat{\sigma}_{k}$, ($k=T,L,A,I$). 
At NLO, the soft-gluon corrections to NLL accuracy in the $\overline{\mathrm{MS}}$ scheme read  \cite{Laenen-Moch}
\begin{eqnarray}
K^{(1)}( s^{\prime },t_{1},u_{1}) &=& 2\left[ \frac{\ln \left( s_{4}/m^{2}\right)
}{s_{4}}\right]_{+}-\left[\frac{1}{s_{4}}\right]_{+}\Biggl[1+\ln \frac{u_{1}}{t_{1}} \nonumber \\
&&-\left( 1-\frac{2C_{F}}{ C_{A}}\right) \left(
1+\mathrm{Re}L_{\beta }\right) +\ln \frac{\mu ^{2}}{m^{2}} \Biggr]   \nonumber \\
&&{}+\delta ( s_{4}) \ln \frac{-u_{1}}{m^{2}} \ln \frac{\mu ^{2}}{m^{2}}. \label{17}
\end{eqnarray}
In Eq.~(\ref{17}), $C_{A}=N_{c}$, $ C_{F}=(N_{c}^{2}-1)/(2N_{c})$, $N_{c}$ is the number of quark colors, and $ L_{\beta }=(1-2m^{2}/s)\{\ln[(1-\beta_{z})/(1+\beta_{z})]+$i$\pi\}$ with $\beta_{z}=\sqrt{1-4m^{2}/s}$. The single-particle inclusive ``plus'' distributions are defined by
\begin{eqnarray}  
\left[\frac{\ln^{l}\left( s_{4}/m^{2}\right) }{s_{4}}\right]_{+}&=&\lim_{\epsilon
\rightarrow 0}\Biggl[\frac{\ln^{l}\left(s_{4}/m^{2}\right) }{s_{4}}\theta (
s_{4}-\epsilon)  \nonumber \\
&&+\frac{1}{l+1}\ln ^{l+1}\frac{\epsilon }{m^{2}}\delta (s_{4})\Biggr].\label{18}
\end{eqnarray}
For any sufficiently regular test function $h(s_{4})$, Eq.~(\ref{18}) implies that
\begin{eqnarray} 
&&\int_{0}^{s_{4}^{\max }}\mathrm{d}s_{4}\,h(s_{4})\left[ \frac{\ln ^{l}\left(
s_{4}/m^{2}\right) }{s_{4}}\right]_{+}  \nonumber \\
&&=\int_{0}^{s_{4}^{\max}}\mathrm{d}s_{4}\left[ h(s_{4})-h(0)\right] \frac{\ln ^{l}\left( s_{4}/m^{2}\right)}{s_{4}}   \nonumber \\
&&~~+\frac{1}{l+1}h(0)\ln ^{l+1}\frac{s_{4}^{\max }}{m^{2}}.  \label{19}
\end{eqnarray}

Standard NLL soft-gluon approximation allows us to determine unambiguously only the singular $s_{4}$ behavior of the cross sections defined by Eq.~(\ref{18}). To fix the $s_{4}$ dependence of the Born-level distributions $B_{k}^{\mathrm{Born}}(s^{\prime},t_{1},u_{1})$ in Eq.~(\ref{16}), we use the method  proposed in \cite{we7} and based on comparison of the soft-gluon predictions with the exact NLO results. According to \cite{we7},
\begin{eqnarray}
B_{k}^{\mathrm{ Born}}( s^{\prime },t_{1},u_{1})&\equiv& s^{\prime 2}\frac{\mathrm{d}\hat{\sigma}^{(0)}_{k,g}}{\mathrm{d}t_{1}}(x_{4}s^{\prime},x_{4}t_{1}), \nonumber \\
x_{4}&=&-\frac{u_{1}}{s^{\prime}+t_{1}}=1-\frac{s_{4}}{s^{\prime}+t_{1}}, \label{20}
\end{eqnarray}
where the leading order GF differential distributions, 
$\frac{\mathrm{d}\hat{\sigma}^{(0)}_{k,g}}{\mathrm{d}t_{1}}(s^{\prime},t_{1})$, are:
\begin{eqnarray}
\frac{\mathrm{d}\hat{\sigma}^{(0)}_{T,g}}{\mathrm{d}t_{1}}(s^{\prime},t_{1})=\pi e_{Q}^{2}\alpha_{\mathrm{em}}\alpha_{s}&&\!\!\!\frac{1}{s^{\prime 2}}\Biggl\{-\frac{t_{1}}{s^{\prime}+t_{1}}-\frac{s^{\prime}+t_{1}}{t_{1}}  \nonumber  \\
+4\left( \frac{s}{s^{\prime}}+\frac{m^{2}s^{\prime }}{t_{1}(s^{\prime}+t_{1})}\right)&&\!\!\!\left[ \frac{Q^{2}}{s^{\prime}}-\frac{s^{\prime}(m^{2}-Q^{2}/2)}{t_{1}(s^{\prime}+t_{1})}\right] \Biggr\}, \nonumber \\
\frac{\mathrm{d}\hat{\sigma}^{(0)}_{L,g}}{\mathrm{d}t_{1}}(s^{\prime},t_{1})=\pi e_{Q}^{2}\alpha_{\mathrm{em}}\alpha_{s}&&\!\!\!\frac{8Q^{2}}{s^{\prime 3}}\left( \frac{s}{s^{\prime }}+\frac{m^{2}s^{\prime }}{t_{1}(s^{\prime}+t_{1})}\right), \label{21} \\
\frac{\mathrm{d}\hat{\sigma}^{(0)}_{A,g}}{\mathrm{d}t_{1}}(s^{\prime},t_{1})=\pi e_{Q}^{2}\alpha_{\mathrm{em}}\alpha_{s}&&\!\!\!\frac{4}{s^{\prime 2}}\left( \frac{s}{s^{\prime}}+\frac{ m^{2}s^{\prime }}{t_{1}(s^{\prime}+t_{1})}\right) \nonumber \\
&&\!\!\!\times\left(\frac{Q^{2}}{s^{\prime }}-\frac{m^{2}s^{\prime}}{t_{1}(s^{\prime}+t_{1})}\right),  \nonumber  \\
\frac{\mathrm{d}\hat{\sigma}^{(0)}_{I,g}}{\mathrm{d}t_{1}}(s^{\prime},t_{1})=\pi e_{Q}^{2}\alpha_{\mathrm{em}}\alpha_{s}&&\!\!\!\frac{4\sqrt{Q^{2}}}{s^{\prime 2}}\!\!\left(\!\! \frac{-t_{1}s(s^{\prime}+t_{1}) }{s^{\prime 2}}-m^{2}\!\!\right)^{\!\!1/2}  \nonumber  \\
\times\frac{s^{\prime}+2t_{1}}{-t_{1}(s^{\prime}+t_{1})}&&\!\!\!\left( 1-\frac{2Q^{2}}{s^{\prime }}+\frac{2m^{2}s^{\prime}}{t_{1}(s^{\prime}+t_{1})}\right).    \nonumber 
\end{eqnarray}

Comparison with the exact NLO results given by Eqs.~(4.7) and (4.8) in Ref.~\cite{LRSN} indicates that the usage of the distributions $B_{k}^{\mathrm{ Born}}(s^{\prime},t_{1},u_{1})$ defined by Eqs.~(\ref{20}) and (\ref{21}) in present paper provides an accurate account of the logarithmic contributions originating from the collinear gluon emission. 
Numerical analysis shows that Eqs. (\ref{20}) and (\ref{21}) render it possible to describe with good accuracy the exact NLO predictions for the functions $\hat{\sigma}^{(1)}_{T}(s^{\prime})$ and $\hat{\sigma}^{(1)}_{L}(s^{\prime})$  near the threshold at relatively low virtualities $Q^{2}\sim m^{2}$  \cite{we7}.\footnote{Note that soft-gluon approximation is unreliable for high $Q^{2} \gg m^{2}$.}
\begin{figure*}
\begin{center}
\begin{tabular}{cc}
\mbox{\epsfig{file=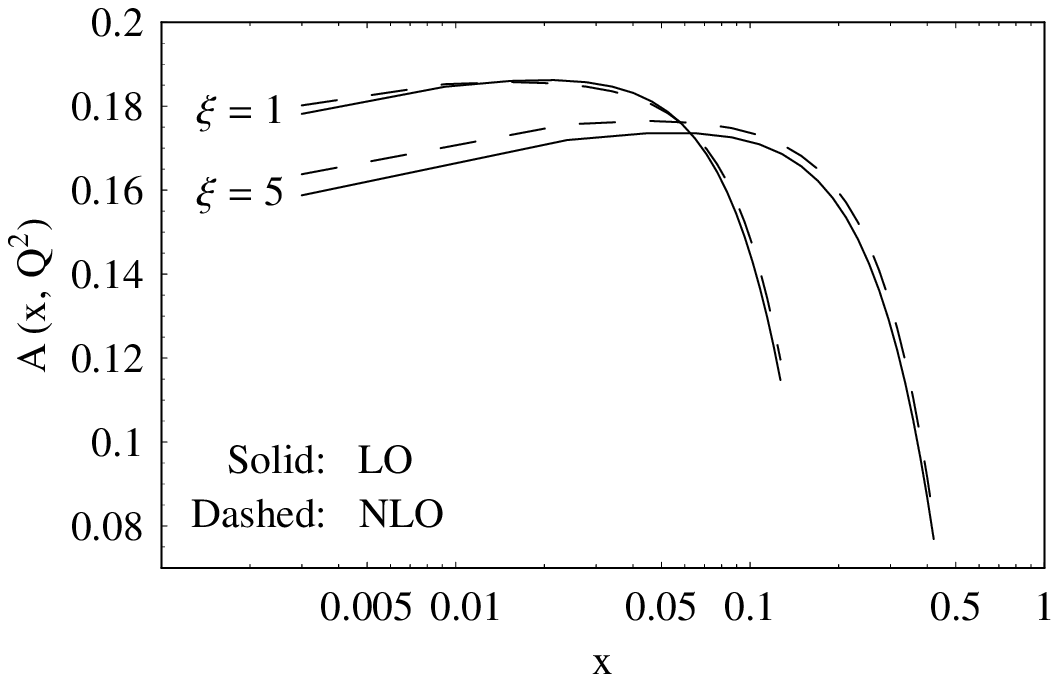,width=230pt}}
& \mbox{\epsfig{file=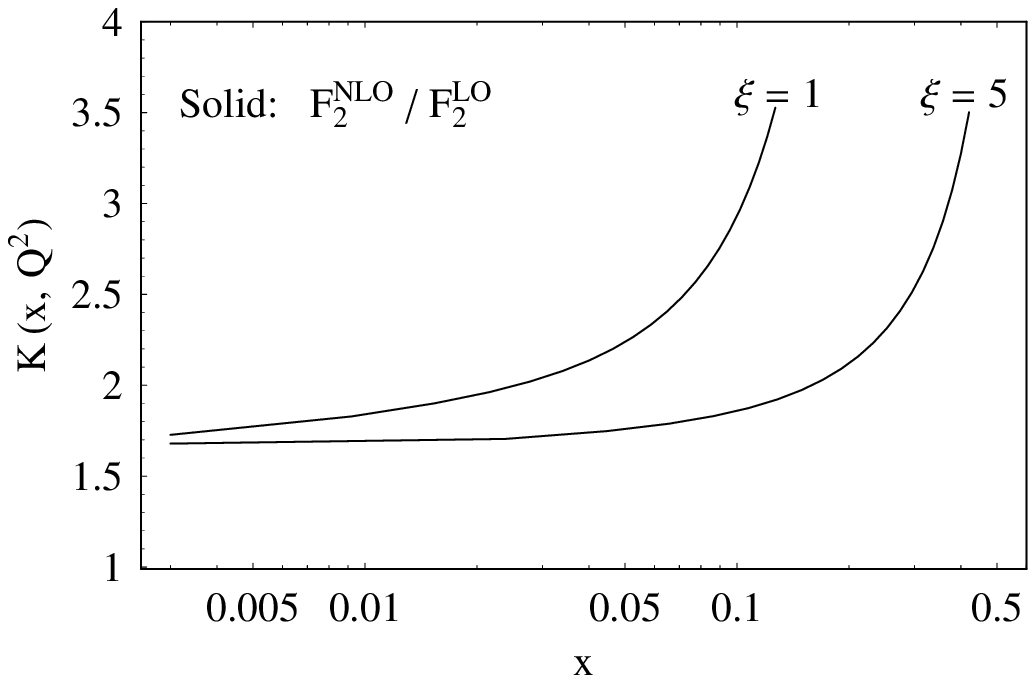,width=230pt}}\\
\end{tabular}
\caption{\label{Fg.4}\small \emph{Left panel:} LO (solid lines) and NLO (dashed lines)  soft-gluon predictions for the $x$ dependence of the azimuthal $\cos(2\varphi)$ asymmetry,
$A(x,Q^{2})=2xF_{A}/F_{2}$, in charm leptoproduction at $\xi=1$ and 5.
\emph{Right panel:} $x$ dependence of the $K$ factor,  $K(x,Q^2)=F_2^{\mathrm{NLO}}/F_2^{\mathrm{LO}}$, at the same values of $\xi$.}
\end{center}
\end{figure*}

Our results for the $x$ distribution of the azimuthal $\cos(2\varphi)$ asymmetry, $A(x,Q^{2})=2xF_{A}/F_{2}$, in charm leptoproduction at fixed values of $\xi$ are presented in the left panel of Fig.~\ref{Fg.4}. For comparison, the $K$ factor,  
$K(x,Q^2)=F_2^{\mathrm{NLO}}/F_2^{\mathrm{LO}}$, for the structure function $F_2$ at the same values of $\xi$ is shown in the right panel of Fig.~\ref{Fg.4}. One
can see that the sizable soft-gluon corrections to the production cross sections affect
the Born predictions for $A(x,Q^2)$ at NLO very little, by a few percent only.

\section{\label{analytic} Analytic LO results for $R(x,Q^2)$ and $A(x,Q^2)$ at low $x$}

Since the ratios $R(x,Q^2)$ and $A(x,Q^2)$ are perturbatively stable, it makes sense to provide the LO hadron-level predictions for these quantities in analytic form. In this Section, we derive compact low-$x$ approximation formulae for the azimuthal $\cos(2\varphi)$ asymmetry and quantity $R_2(x,Q^2)$ closely related to the Callan-Gross ratio $R(x,Q^2)$,
\begin{equation}  \label{22}
R_{2}(x,Q^{2})=2x\frac{F_{L}}{F_{2}}(x,Q^2)=\frac{R(x,Q^{2})}{1+R(x,Q^{2})}.
\end{equation}
We will see below that our obtained results may be useful in the extraction of the structure functions $F_k$ ($k=2,L,A,I$) from measurements of the reduced cross sections. 

To obtain the hadron-level predictions, we convolute the LO partonic cross sections given by Eqs.~(\ref{8}) with the low-$x$ asymptotics of the gluon PDF:
\begin{equation} \label{23}
g(x,Q^2)\stackrel{x\to 0}{\longrightarrow}\frac{1}{x^{1+\delta}}.
\end{equation}

The value of $\delta$ in Eq.~(\ref{23}) is a matter of discussion. The simplest choice,
$\delta =0$, leads to a non-singular behavior of the structure functions for $x\to 0$.\footnote{The LO predictions for the Callan-Gross ratio in the case of $\delta =0$ were  studied in Ref.~\cite{kotikov}.}
Another extreme value, $\delta =1/2$, historically originates from the BFKL resummation 
of the leading powers of $\ln(1/x)$ \cite{BFKL1,BFKL2,BFKL3}. In reality, $\delta$ is a function of $Q^2$. Theoretically, the $Q^2$ dependence of $\delta$ is calculated using the DGLAP evolution equations \cite{DGLAP1,DGLAP2,DGLAP3}.

We have derived the analytic low-$x$ formulae for the ratios $A^{(\delta )}(Q^2)\equiv A^{(\delta)}(x\to 0,Q^2)$ and $R_{2}^{(\delta )}(Q^2)\equiv R^{(\delta)}_2(x\to 0,Q^2)$ with arbitrary values of $\delta$ in terms of the Gauss hypergeometric function. Our results have the following form:\\

\begin{strip}
\begin{picture}(240,10)
\put(0,10){\line(1,0){240}}
\end{picture}
\begin{equation} \label{29}
\qquad \qquad \qquad A^{(\delta )}(Q^2)=2\frac{\frac{2+\delta +2\lambda}{3+\delta }\mathrm {\Phi}
\left( 1+\delta ,\frac{1}{1+4\lambda }\right) -\left( 1+4\lambda \right)
\mathrm{\Phi} \left( 2+\delta ,\frac{1}{1+4\lambda }\right) }{\left[ 1+\frac{%
\delta \left( 1-\delta ^{2}\right) }{\left( 2+\delta \right) \left( 3+\delta \right)
}\right] \mathrm{\Phi} \left( \delta ,\frac{1}{1+4\lambda }\right) -\left( 1+4\lambda \right)
\left( 4-\delta -\frac{10}{3+\delta }\right) \mathrm{\Phi} \left( 1+\delta ,\frac{1}{1+4\lambda
}\right) },
\end{equation}
\begin{equation} \label{24}
\qquad \qquad \qquad R_{2}^{(\delta )}(Q^2)=4\frac{\frac{2+\delta }{3+\delta }\mathrm{\Phi}
\left( 1+\delta ,\frac{1}{1+4\lambda }\right) -\left( 1+4\lambda \right)
\mathrm{\Phi} \left( 2+\delta ,\frac{1}{1+4\lambda }\right) }{\left[ 1+\frac{%
\delta \left( 1-\delta ^{2}\right) }{\left( 2+\delta \right) \left( 3+\delta \right)
}\right] \mathrm{\Phi} \left( \delta ,\frac{1}{1+4\lambda }\right) -\left( 1+4\lambda \right)
\left( 4-\delta -\frac{10}{3+\delta }\right) \mathrm{\Phi} \left( 1+\delta ,\frac{1}{1+4\lambda
}\right) },
\end{equation}
\begin{flushright}
\begin{picture}(240,10)
\put(0,-10){\line(1,0){240}}
\end{picture}
\end{flushright}
\end{strip}

\noindent where $\lambda=m^2/Q^2$ and the function $\mathrm{\Phi}\left(r,z\right)$ is 
\begin{eqnarray} 
\mathrm{\Phi} \left( r,z\right)&=&\frac{z^{1+r}}{1+r}\,\frac{\mathrm{\Gamma} \left( 1/2\right) \mathrm{\Gamma} \left(1+r\right) }{\mathrm{\Gamma} \left( 3/2+r\right)} \nonumber \\
&&\times\,{}_{2}F_{1}\left(\frac{1}{2},1+r,\frac{3}{2}+r;z\right).\label{25}
\end{eqnarray}
The hypergeometric function ${}_{2}F_{1}(a,b;c;z)$ has the following series expansion:
\begin{eqnarray}
{}_{2}F_{1}\left( a,b,c;z\right)&=&\frac{\mathrm{\Gamma} \left( c\right) }{\mathrm{\Gamma} \left( a\right)\mathrm{\Gamma} \left( b\right)}  \nonumber \\
&&\times\sum\limits_{n=0}^{\infty }\frac{\mathrm{\Gamma}
\left( a+n\right) \mathrm{\Gamma} \left( b+n\right) }{\mathrm{\Gamma} \left( c+n\right) }\frac{%
z^{n}}{n!}.  \label{26}
\end{eqnarray}

\begin{figure*}
\begin{center}
\begin{tabular}{cc}
\mbox{\epsfig{file=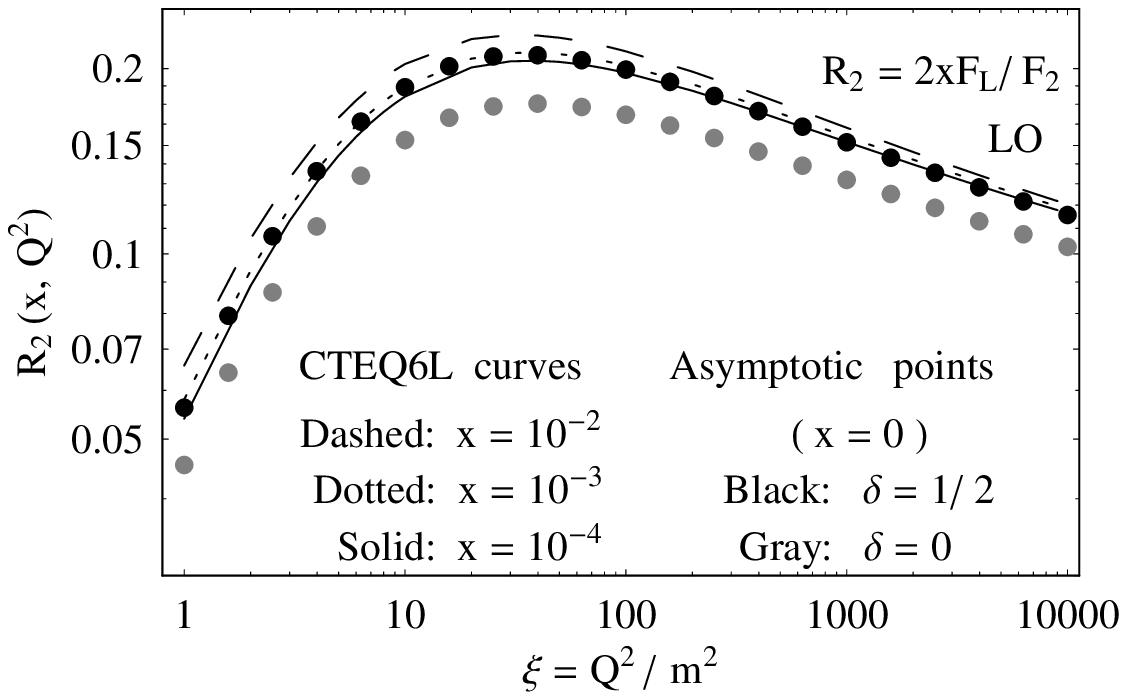,width=230pt}}
& \mbox{\epsfig{file=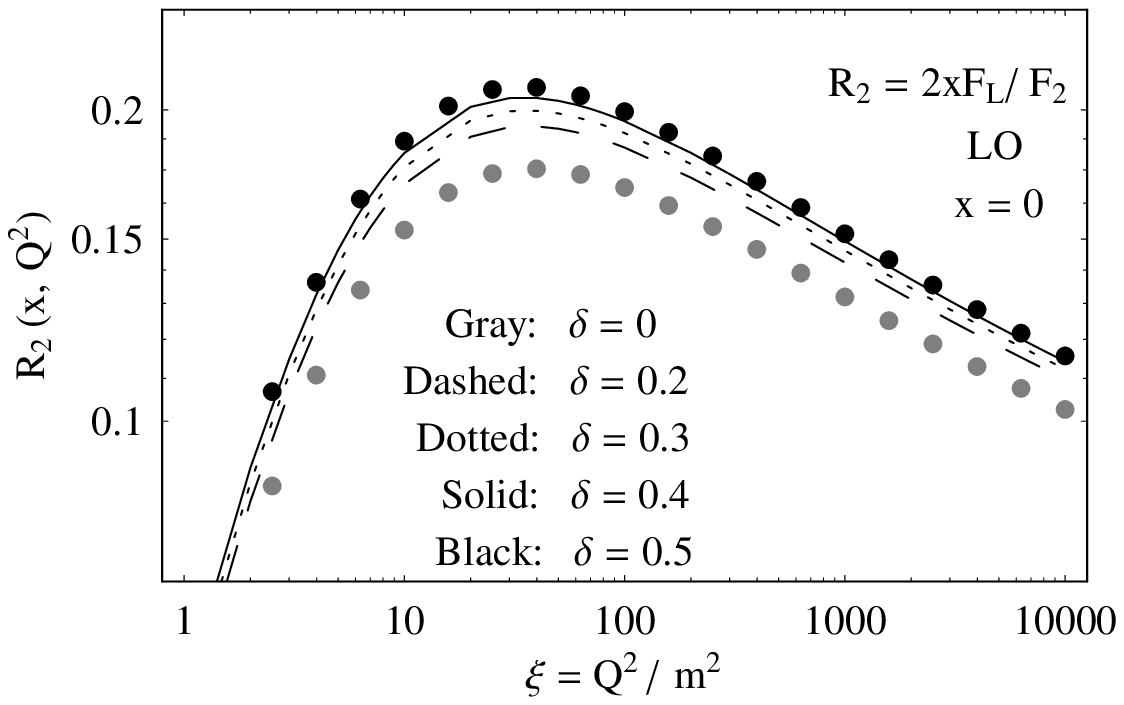,width=230pt}}\\
\end{tabular}
 \caption{\label{Fg.5}\small LO low-$x$ predictions for the ratio $R_2(x,Q^2)=2xF_L/F_2$
 in charm leptoproduction. \emph{Left panel:} Asymptotic ratios $R^{(0)}_2(Q^2)$
 (gray points) and $R^{(1/2)}_2(Q^2)$ (black points), as well as CTEQ6L
 predictions for $R_2(x,Q^2)$ at $x=10^{-2}$, $10^{-3}$ and $10^{-4}$.
 \emph{Right panel:}
 Asymptotic ratio $R^{(\delta )}_2(Q^2)$ at
 $\delta=0$, 0.2, 0.3, 0.4 and 0.5.}
\end{center}
\end{figure*}

In Fig.~\ref{Fg.5}, we investigate the result (\ref{24}) for  $R_{2}^{(\delta )}(Q^2)$. The left panel of Fig.~\ref{Fg.5} shows the ratio $R_{2}^{(\delta )}(Q^2)$ as functions of $\xi$ for two  extreme cases, $\delta =0$ and $1/2$. One can see that the difference between these quantities varies slowly from $20\%$ at low $Q^2$ to $10\%$ at high $Q^2$. For comparison, the LO results for $R_2(x,Q^2)$ are also shown at several values of $x$. In calculations, the CTEQ6L gluon PDF \cite{CTEQ6} was  used. We observe that, for $x\to 0$, the
CTEQ6L predictions converge to the function $R^{(1/2)}_2(Q^2)$ practically in the entire
region of $Q^2$. We have verified that the similar situation takes also place for other 
CTEQ PDF versions \cite{CT10,CT14}.
In the right panel of Fig.~\ref{Fg.5}, the $\delta$ dependence of the asymptotic ratio
$R^{(\delta)}_2(Q^2)$ is investigated. One can see that the ratio $R^{(\delta)}_2(Q^2)$
rapidly converges to the function $R^{(1/2)}_2(Q^2)$ for $\delta > 0.2$. In particular,
the relative difference between $R^{(0.5)}_2(Q^2)$ and $R^{(0.3)}_2(Q^2)$ varies slowly
from $6\%$ at low $Q^2$ to $2\%$ at high $Q^2$.

\begin{figure*}
\begin{center}
\begin{tabular}{cc}
\mbox{\epsfig{file=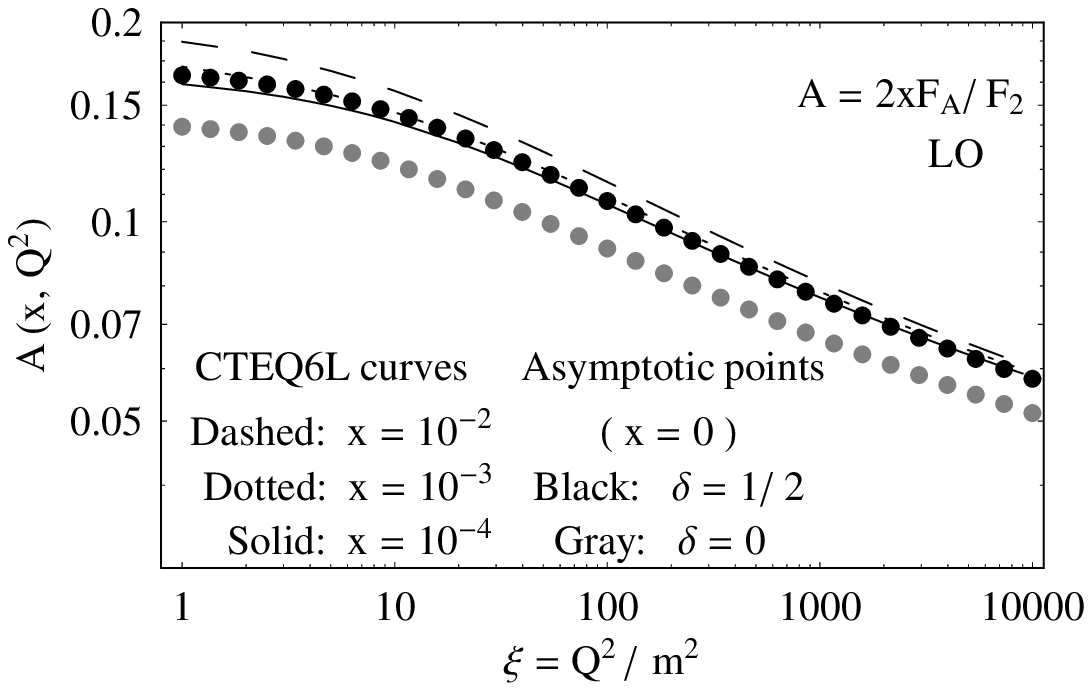,width=230pt}}
& \mbox{\epsfig{file=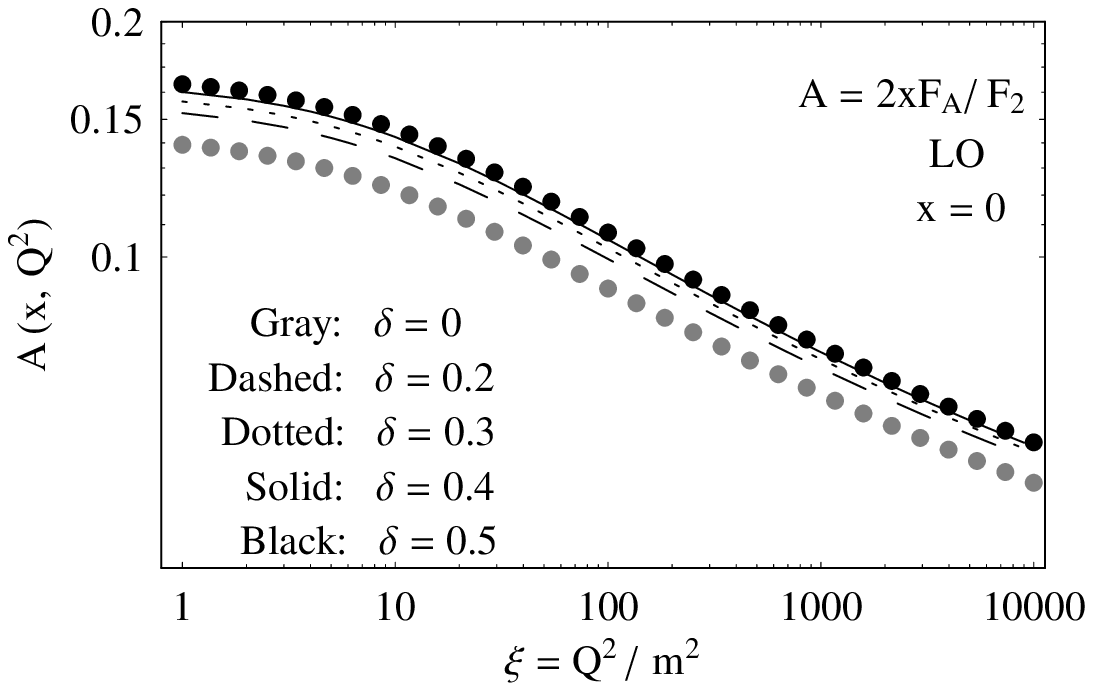,width=230pt}}\\
\end{tabular}
 \caption{\label{Fg.6}\small LO low-$x$ predictions for the ratio $A(x,Q^2)=2xF_A/F_2$
 in charm leptoproduction. \emph{Left panel:} Asymptotic ratios $A^{(0)}(Q^2)$
 (gray points) and $A^{(1/2)}(Q^2)$ (black points), as well as CTEQ6L
 predictions for $A(x,Q^2)$ at $x=10^{-2}$, $10^{-3}$ and $10^{-4}$.
 \emph{Right panel:}
 Asymptotic ratio $A^{(\delta )}(Q^2)$ at
 $\delta=0$, 0.2, 0.3, 0.4 and 0.5.}
\end{center}
\end{figure*}

Our analysis presented in Fig.~\ref{Fg.6} shows that the quantity $A^{(\delta )}(Q^2)$ defined by Eq.~(\ref{29}) has the properties very similar to the ones demonstrated by the ratio $R^{(\delta)}_2(Q^2)$. In particular, one can see from Fig.~\ref{Fg.6} that the hadron-level predictions for $A^{(\delta )}(Q^2)$ depend weakly on $\delta$ practically in the entire region of $Q^2$ for $\delta > 0.2$. 

As mentioned above, the $Q^2$ dependence of the parameter $\delta$ is determined with the 
help of the DGLAP evolution. However, our analysis shows that hadron-level predictions for both $A^{(\delta)}(x\to 0,Q^2)$ and $R^{(\delta)}_2(x\to 0,Q^2)$ depend weakly on $\delta$ practically in the entire region of $Q^2$ for $0.2< \delta < 0.9$. For this reason, it makes sense to consider the ratios $A^{(\delta )}(Q^2)$ and $R^{(\delta)}_2(Q^2)$ in particular case of $\delta = 1/2$. The results are:\\

\begin{strip}
\begin{picture}(240,10)
\put(0,10){\line(1,0){240}}
\end{picture}
\begin{equation} \label{27a}
\qquad \qquad \qquad A^{(1/2)}(Q^2)=12\frac{(1+ 8\lambda) E(1/(1 + 4\lambda)) - 8\lambda K(1/(1 + 4\lambda))}{ \left( -37 + 72\lambda
\right)E(1/(1 + 4\lambda)) + 2\left( 23 - 36\lambda
\right)K(1/(1 + 4\lambda)) },
\end{equation}
\begin{equation} \label{27}
\qquad \qquad \qquad R^{(1/2)}_2(Q^2)=\frac{8}{1 + 4\lambda}\,\frac{ \left[ 3 + 4\lambda \left( 13
+ 32\lambda \right) \right] E(1/(1 + 4\lambda)) - 4\lambda \left( 9 +
32\lambda  \right) K(1/(1 + 4\lambda)) }{ \left( -37 + 72\lambda
\right)E(1/(1 + 4\lambda)) + 2\left( 23 - 36\lambda
\right)K(1/(1 + 4\lambda)) },
\end{equation}
\begin{flushright}
\begin{picture}(240,10)
\put(0,-10){\line(1,0){240}}
\end{picture}
\end{flushright}
\end{strip}

\noindent where the functions $K(y)$ and $E(y)$ are the complete elliptic integrals of the first and second kinds defined as
\begin{equation} \label{28}
K(y)=\int\limits_0^1\!\!\!\frac{{\mathrm d}t}{\sqrt{(1-t^2)(1-yt^2)}},\;
E(y)=\int\limits_0^1\!{\mathrm d}t \sqrt{\frac{1-yt^2}{1-t^2}}.
\end{equation}

One can see from Figs.~\ref{Fg.5} and \ref{Fg.6} that our simple formulae (\ref{27}) and (\ref{27a}) with $\delta =1/2$ (i.e., without any evolution) describes with good accuracy the low-$x$ CTEQ results for $R_2(x,Q^2)$ and $A(x,Q^2)$. We conclude that the hadron-level predictions for both $R_2(x\to 0,Q^2)$ and $A(x\to 0,Q^2)$ are stable not only under the NLO corrections to the partonic cross sections, but also under the DGLAP evolution of the gluon PDF.

Let us now discuss how the obtained analytic results may be used in the extraction of the structure functions  $F_k$  ($k=2,L,A,I$) from experimental data. Usually, it is the so-called "reduced cross section", $\tilde{\sigma}(x,Q^{2})$, that can directly be measured in DIS experiments:
\begin{eqnarray}
\tilde{\sigma}(x,Q^{2})&=&\frac{1}{1+(1-y)^2}\frac{xQ^4}{2\pi\alpha^{2}_{\mathrm{em}}}\frac{\mathrm{d}^{2}\sigma_{lN}}{\mathrm{d}x\mathrm{d}Q^{2}} \nonumber \\
&=&F_{2}(x,Q^{2})-\frac{2xy^{2}}{1+(1-y)^2}F_{L}(x,Q^{2}) \label{30} \\
&=&F_{2}(x,Q^{2})\left[1-\frac{y^{2}}{1+(1-y)^2}R_{2}(x,Q^{2})\right]\!\!. \label{31}
\end{eqnarray} 
In earlier HERA analyses of charm and bottom electroproduction, the corresponding longitudinal structure functions were taken to be zero for simplicity. In this case, $\tilde{\sigma}(x,Q^{2})=F_{2}(x,Q^{2})$. In later papers, the structure function $F_{2}(x,Q^2)$ is evaluated from the reduced cross section (\ref{30}) where the longitudinal structure function $F_{L}(x,Q^2)$ is estimated from the NLO QCD expectations. Instead of this rather cumbersome procedure, we propose to use the expression (\ref{31}) with the quantity $R_{2}(x,Q^2)$ defined by the analytic LO expressions (\ref{24}) or (\ref{27}). This simplifies the extraction of $F_{2}(x,Q^2)$ from measurements of $\tilde{\sigma}(x,Q^{2})$ but does not affect the accuracy of the result in practice because of perturbative stability of the ratio $R_{2}(x,Q^2)$. 

In Table~\ref{tab1} and \ref{tab2}, we compare the results of our analysis of the last HERA data on the charm and bottom electroproduction with the NLO values, $F_2(\mathrm{NLO})$, obtained by the H1 collabotation \cite{H1HERA1}. One can see that the LO Eq.~(\ref{27}) reproduce the NLO H1  results for $F_2^c(x,Q^2)$ and $F_2^b(x,Q^2)$ with an accuracy better than 1\%.
\begin{table*}
\caption{\label{tab1} Values of $F_2^c(x,Q^2)$ extracted from the HERA measurements of
$\tilde{\sigma}^c(x,Q^{2})$ for various values of $Q^2$ and $x$. The NLO H1 results \cite{H1HERA1} are compared with the LO predictions corresponding to
the case of $\delta =0.5$.}
\begin{center}
\begin{tabular}{||ccc||cc||cc||}
\hline
 $\quad Q^2 \quad$ & $x$ & ~$\quad y\quad$~ & ~$\quad \tilde{\sigma}^c \quad$~ &
  ~Error~ &
$\qquad F_2^c$(NLO)$\qquad$ & $ F_2^c$(LO)  \\
 (GeV$^2$) &  &  &  & (\%) & H1 & $\delta=0.5$ \\
\hline
  \hline
    5.0 & 0.00020 & 0.246 & 0.148 & 17.6 & $0.149\pm0.026$ & $0.149\pm0.026$ \\
    8.5 & 0.00050 & 0.167 & 0.176 & 14.8 & $0.176\pm0.026$ & $0.176\pm0.026$ \\
    8.5 & 0.00032 & 0.262 & 0.186 & 15.5 & $0.187\pm0.029$ & $0.187\pm0.029$ \\
   12.0 & 0.00130 & 0.091 & 0.150 & 18.7 & $0.150\pm0.028$ & $0.150\pm0.028$ \\
   12.0 & 0.00080 & 0.148 & 0.177 & 15.9 & $0.177\pm0.028$ & $0.177\pm0.028$ \\
   12.0 & 0.00050 & 0.236 & 0.240 & 11.2 & $0.242\pm0.027$ & $0.241\pm0.027$ \\
   12.0 & 0.00032 & 0.369 & 0.273 & 13.8 & $0.277\pm0.038$ & $0.277\pm0.038$ \\
   20.0 & 0.00200 & 0.098 & 0.187 & 12.7 & $0.188\pm0.023$ & $0.187\pm0.024$ \\
   20.0 & 0.00130 & 0.151 & 0.219 & 11.9 & $0.219\pm0.026$ & $0.220\pm0.026$ \\
   20.0 & 0.00080 & 0.246 & 0.274 & 10.2 & $0.276\pm0.028$ & $0.276\pm0.028$ \\
   20.0 & 0.00050 & 0.394 & 0.281 & 13.8 & $0.287\pm0.040$ & $0.287\pm0.040$ \\
   35.0 & 0.00320 & 0.108 & 0.200 & 12.7 & $0.200\pm0.025$ & $0.200\pm0.025$ \\
   35.0 & 0.00200 & 0.172 & 0.220 & 11.8 & $0.220\pm0.026$ & $0.221\pm0.026$ \\
   35.0 & 0.00130 & 0.265 & 0.295 &  9.7 & $0.297\pm0.029$ & $0.298\pm0.029$ \\
   35.0 & 0.00080 & 0.431 & 0.349 & 12.7 & $0.360\pm0.046$ & $0.359\pm0.046$ \\
   60.0 & 0.00500 & 0.118 & 0.198 & 10.8 & $0.199\pm0.021$ & $0.198\pm0.021$ \\
   60.0 & 0.00320 & 0.185 & 0.263 &  8.4 & $0.264\pm0.022$ & $0.264\pm0.022$ \\
   60.0 & 0.00200 & 0.295 & 0.335 &  8.8 & $0.339\pm0.030$ & $0.339\pm0.030$ \\
   60.0 & 0.00130 & 0.454 & 0.296 & 15.1 & $0.307\pm0.046$ & $0.306\pm0.046$ \\
  120.0 & 0.01300 & 0.091 & 0.133 & 14.1 & $0.133\pm0.019$ & $0.133\pm0.019$ \\
  120.0 & 0.00500 & 0.236 & 0.218 & 11.1 & $0.220\pm0.024$ & $0.220\pm0.024$ \\
  120.0 & 0.00200 & 0.591 & 0.351 & 12.8 & $0.375\pm0.048$ & $0.374\pm0.048$ \\
  200.0 & 0.01300 & 0.151 & 0.160 & 11.9 & $0.160\pm0.019$ & $0.160\pm0.019$ \\
  200.0 & 0.00500 & 0.394 & 0.237 & 13.5 & $0.243\pm0.033$ & $0.242\pm0.033$ \\
  300.0 & 0.02000 & 0.148 & 0.117 & 18.5 & $0.117\pm0.022$ & $0.117\pm0.022$ \\
  300.0 & 0.00800 & 0.369 & 0.273 & 12.7 & $0.278\pm0.035$ & $0.278\pm0.035$ \\
  650.0 & 0.03200 & 0.200 & 0.084 & 30.9 & $0.085\pm0.026$ & $0.084\pm0.026$ \\
  650.0 & 0.01300 & 0.492 & 0.195 & 16.2 & $0.203\pm0.033$ & $0.202\pm0.033$ \\
 2000.0 & 0.05000 & 0.394 & 0.059 & 36.4 & $0.060\pm0.022$ & $0.060\pm0.022$ \\
\hline
\end{tabular}
\end{center}
\end{table*} 

\begin{table*}
\caption{\label{tab2} Values of $F_2^b(x,Q^2)$ extracted from the HERA measurements of
$\tilde{\sigma}^b(x,Q^{2})$ for various values of $Q^2$ and  $x$. The NLO H1 results \cite{H1HERA1} are compared with the LO predictions corresponding to
the case of $\delta =0.5$.}
\begin{center}
\begin{tabular}{||ccc||cc||cc||}
\hline
 $\quad Q^2 \quad$ & $x$ & ~$\quad y\quad$~ & ~$\quad \tilde{\sigma}^b \quad$~ &
  ~Error~ &
$\qquad F_2^b$(NLO)$\qquad$ & $ F_2^b$(LO)  \\
 (GeV$^2$) &  &  &  & (\%) & H1 & $\delta=0.5$ \\
\hline
  \hline
 5. & 0.00020 & 0.246 & 0.00244 & 46.1 & $0.00244\pm0.00112$ & $0.00244\pm0.00113$ \\
 12. & 0.00032 & 0.369 & 0.00487 & 31.8 & $0.00490\pm0.00156$ & $0.00489\pm0.00156$ \\
 12. & 0.00080 & 0.148 & 0.00247 & 43.5 & $0.00248\pm0.00108$ & $0.00247\pm0.00108$ \\
 25. & 0.00050 & 0.492 & 0.01189 & 25.1 & $0.01206\pm0.00303$ & $0.01203\pm0.00302$ \\
 25. & 0.00130 & 0.189 & 0.00586 & 34.1 & $0.00587\pm0.00200$ & $0.00587\pm0.00200$ \\
 60. & 0.00130 & 0.454 & 0.01928 & 25. & $0.01969\pm0.00492$ & $0.01962\pm0.00490$ \\
 60. & 0.00500 & 0.118 & 0.00964 & 32.6 & $0.00965\pm0.00315$ & $0.00965\pm0.00315$ \\
 200. & 0.00500 & 0.394 & 0.02365 & 23.2 & $0.02422\pm0.00562$ & $0.02415\pm0.00560$ \\
 200. & 0.01300 & 0.151 & 0.01139 & 34.4 & $0.01142\pm0.00393$ & $0.01142\pm0.00393$ \\
 650. & 0.01300 & 0.492 & 0.01331 & 34.7 & $0.01394\pm0.00484$ & $0.01388\pm0.00481$ \\
 650. & 0.03200 & 0.200 & 0.01018 & 30.1 & $0.01024\pm0.00308$ & $0.01023\pm0.00308$ \\
 2000. & 0.05000 & 0.394 & 0.00499 & 61.1 & $0.00511\pm0.0031$ & $0.00511\pm0.00312$ \\
\hline
\end{tabular}
\end{center}
\end{table*} 

High accuracy of our LO approach is explained as follows. 
One can see from Eq.(32) that the LO corrections to the extracted function
$F_2(x,Q^2)$ due to the non-zero value of $R_2(x,Q^2)$ cannot
exceed $30\%$ because the ratio $R_2(x,Q^2)$ is itself less than
0.3 practically in the entire region of the variables $x$ and $Q^2$.
For this reason, the NLO corrections to $R_2(x,Q^2)$, having
a relative size of the order of $10\%$, cannot affect the value
of $F_2(x,Q^2)$ by more than $3\%$. In reality, the effect of radiative
corrections to $R_2(x,Q^2)$ on the extracted values of
$F_2(x,Q^2)$ is less than $1\%$ since $y\ll 1$ in most of the experimentally
accessible kinematic range. 

Taking into account that typical experimental errors are of about (10--20)$\%$, 
we conclude that our analytic predictions for $R_2(x,Q^2)$ and $A(x,Q^2)$ will be useful in extraction of the structure functions from presently available and future data. 

The structure functions $F_A$ and $F_I$ can be extracted from the $\varphi$-dependent DIS cross section,
\begin{eqnarray}
\frac{\mathrm{d}^{3}\sigma_{lN}}{\mathrm{d}x\mathrm{d}Q^{2}\mathrm{d}\varphi}
&=&\frac{2\alpha^{2}_{em}y^2}{Q^4(1-\varepsilon)}\Biggl[\frac{1}{2x} F_{2}( x,Q^{2})- (1-\varepsilon) F_{L}(x,Q^{2}) \Bigr. \nonumber \\
&&+\varepsilon  F_{A}( x,Q^{2})\cos 2\varphi  \nonumber \\
&&+\Bigl. 2\sqrt{\varepsilon(1+\varepsilon)} F_{I}( x,Q^{2})\cos \varphi\Biggr], \label{32}
\end{eqnarray}
where $\varepsilon=\frac{2(1-y)}{1+(1-y)^2}$. 
For this purpose, one should measure the first moments of the $\cos(\varphi)$ and $\cos(2\varphi)$ distributions defined as
\begin{equation}  \label{33}
\langle \cos n\varphi \rangle (x,Q^{2})= \frac{\int_{0}^{2\pi }\mathrm{d}\varphi \cos
n\varphi  \frac{\mathrm{d}^{3}\sigma _{lN}} {\mathrm{d}x\mathrm{d}Q^{2}\mathrm{d}\varphi } (x,Q^{2},\varphi ) }{\int_{0}^{2\pi }\mathrm{d}\varphi \frac{\mathrm{d}^{3}\sigma_{lN}} {\mathrm{d}x\mathrm{d}Q^{2}\mathrm{d}\varphi } (x,Q^{2},\varphi ) }. 
\end{equation}
Using Eq.~(\ref{32}), we obtain:
\begin{eqnarray} 
\langle \cos 2\varphi \rangle(x,Q^{2})&=&\frac{1}{2}\frac{\varepsilon A(x,Q^{2})}{1-(1-\varepsilon)R_2(x,Q^{2})},  \nonumber \\
A(x,Q^{2})&=&2x\frac{F_{A}}{F_{2}}(x,Q^{2}),\label{34}
\end{eqnarray}
and
\begin{eqnarray} 
\langle \cos \varphi \rangle(x,Q^{2})&=&\frac{\sqrt{\varepsilon (1+\varepsilon)} A_I(x,Q^{2})}{1-(1-\varepsilon)R_2(x,Q^{2})},  \nonumber \\
A_I(x,Q^{2})&=&2x\frac{F_{I}}{F_{2}}(x,Q^{2}).\label{35}
\end{eqnarray}

One can see from Eqs.~(\ref{34}) and (\ref{35}) that, using the perturbatively stable predictions (\ref{24}) for $R_2(x,Q^{2})$, we will be able to determine the structure functions $F_A(x,Q^{2})$ and $F_I(x,Q^{2})$ from future data on the moments $\langle\cos 2\varphi\rangle$ and $\langle\cos \varphi\rangle$. On the other hand, according to Eq.~(\ref{34}), the analytic results (\ref{24}) and (\ref{29}) for the quantities $R_2(x,Q^{2})$ and $A(x,Q^{2})$ provide us with the perturbatively stable predictions for $\langle\cos 2\varphi\rangle$ which may be directly tested in experiment.

So, our obtained analytic and perturbatively stable predictions for the ratios $R_2(x,Q^{2})$ and $A(x,Q^{2})$ will simplify both the extraction of structure functions from measurements of the $\varphi$-dependent cross section (\ref{32}) and test of self-consistency of the extraction procedure.  

\section{\label{resum} Resummation of the Mass Logarithms}

In this Section, we discuss the properties of the quantities $R(x,Q^2)$ and $A(x,Q^2)$ within the variable-flavor-number scheme (VFNS) \cite{ACOT,Collins}.  The VFNS is an 
approach alternative to the traditional fixed-flavor-number scheme (FFNS) where only light degrees of freedom ($u,d,s$ and $g$) are considered as active. Within the VFNS, the
mass logarithms of the type $\alpha_{s}\ln\left( Q^{2}/m^{2}\right)$ are resummed through the all orders into a heavy quark density which evolves with $Q^{2}$ according to the standard DGLAP \cite{DGLAP1,DGLAP2,DGLAP3} evolution equations. Hence this approach introduces the parton distribution functions (PDF) for the heavy quarks and changes the number of active flavors by one unit when a heavy quark threshold is crossed. 

At leading order, ${\cal O}(\alpha_{s}^0)$, the only photon-quark scattering (QS) subprocess within the VFNS is 
\begin{equation}
\gamma ^{*}(q)+Q(k_{Q})\rightarrow Q(p_{Q}).  \label{36}
\end{equation}
Corresponding Feynman diagram is depicted in Fig.~\ref{Fg.2}\emph{b}.

The  ${\cal O}(\alpha_{s}^0)$ $\gamma ^{*}Q$ cross sections, $\hat{\sigma}_{k,\mathrm{Q}}^{(0)}(z,\lambda)$,  are:
\begin{eqnarray}
\hat{\sigma}_{2,\mathrm{Q}}^{(0)}(z,\lambda)&=&\hat{\sigma}_{B}(z)\sqrt{1+4\lambda z^{2}}\,
\delta(1-z), \nonumber \\
\hat{\sigma}_{L,\mathrm{Q}}^{(0)}(z,\lambda)&=&\hat{\sigma}_{B}(z)\frac{4\lambda z^{2}}
{\sqrt{1+4\lambda z^{2}}}\,\delta(1-z), \label{37} \\
\hat{\sigma}_{A,\mathrm{Q}}^{(0)}(z,\lambda)&=&\hat{\sigma}_{I,\mathrm{Q}}^{(0)}(z,\lambda)=0, \nonumber 
\end{eqnarray}
with $z=Q^{2}/(2q\cdot k_{Q})$ and $\hat{\sigma}_{B}(z)=(2\pi)^2e_{Q}^{2}\alpha_{\mathrm{em}}\,z/Q^{2}$.

Within the VFNS, the mass logarithms of the type $\alpha_s^n\ln^n(Q^{2}/m^{2})$, which dominate
the production cross sections at high energies, $Q^{2}\rightarrow \infty$, are resummed via  the renormalization group equations. In practice, the resummation procedure consists of two steps. First, the mass logarithms have to be subtracted from the fixed order predictions for the partonic cross sections in such a way that, in the limit $Q^{2}\rightarrow \infty$, the well known massless $\overline{\text{MS}}$ coefficient functions are recovered. Instead, a heavy-quark  density in the hadron, $h(x,Q^{2})$, has to be introduced. This density obeys the usual massless NLO DGLAP evolution equation with the boundary condition $h(x,Q^{2}=Q_{0}^2)=0$ where $Q_{0}^2\sim m^{2}$.

Within the VFNS, the treatment of heavy quarks depends on the values chosen for $Q^{2}$. At low
$Q^{2}<Q_{0}^2$, the production cross sections are described by the light parton contributions
($u,d,s$ and $g$). The heavy-flavor production is dominated by the GF process and its higher order QCD corrections. At high $Q^{2}\gg m^{2}$, the heavy quark is treated in the same way as the other light quarks and it is represented by a heavy-quark parton density in the hadron. In the intermediate scale region one has to make a smooth connection between the two different prescriptions.

Strictly speaking, the perturbative heavy-flavor density is well defined at high $Q^2\gg m^2$ but does not have a clean interpretation at low $Q^2$. Since the heavy-quark distribution originates from resummation of the mass logarithms of the type $\alpha_s^n\ln^n (Q^{2}/m^{2})$, it is usually assumed that the corresponding PDF vanishes with these logarithms, i.e. for $Q^{2}<Q_{0}^2\sim m^{2}$. On the other hand, the threshold constraint $W^2=(q+p)^2=Q^2(1/x-1)>4m^2$ implies that $Q_0$ is not a constant but "live" function of $x$. To avoid this problem, several solutions have been proposed. (For a review, see Ref.~\cite{sacot12-2}.) 

In our analysis, the so-called ACOT($\chi$) scheme \cite{chi} is used. According to the ACOT($\chi$) prescription, the lowest order, ${\cal O}(\alpha_{s})$, hadron-level cross section for charm production is
\begin{eqnarray}
&&\sigma^{(\mathrm{ACOT})}_{2}(x,\lambda)=\hat{\sigma}_{B}(x)c_{+}(\chi,\mu_{F})+\int\limits_{\chi}^{1}\text{d}z\,g(z,\mu_{F})  \nonumber \\
&&\times\!\!\left[\hat{\sigma}_{2,\mathrm{g}}^{(0)}
\!\left(x/z,\lambda\right)-\frac{\alpha_{s}}{\pi}\ln\frac{\mu_{F}^{2}}{m^{2}}\;\hat{\sigma}_{B}\left(x/z\right)P^{(0)}_{g\rightarrow c}\left(\chi/z\right)\right]\!\!. \label{38}  
\end{eqnarray}
In Eq.~(\ref{38}),
\begin{equation}
\chi=x(1+4\lambda), \label{38a}
\end{equation}
$P^{(0)}_{g\rightarrow c}$ is the LO gluon-quark splitting function, $P^{(0)}_{g\rightarrow c}(\zeta)=\left.\left[(1-\zeta)^{2}+\zeta^{2}\right]\right/2$, $c_{+}(\zeta,\mu_{F})=c(\zeta,\mu_{F})+\bar{c}(\zeta,\mu_{F})$, and the ${\cal O}(\alpha_{s})$ photon-gluon fusion cross section $\hat{\sigma}_{2,\mathrm{g}}^{(0)}$ is given by Eq.~(\ref{8}). 

One can see from Eqs.~(\ref{8}) that the longitudinal and azimuth-dependent cross sections, 
$\hat{\sigma}_{L,\mathrm{g}}^{(0)}$ and $\hat{\sigma}_{A,\mathrm{g}}^{(0)}$, are infra-red safe; the contributions of the potentially large logarithms of the type $\ln (Q^{2}/m^{2})$ to these quantities vanish for $\lambda \to 0$. For this reason, the ${\cal O}(\alpha_{s})$ hadron-level longitudinal and azimuth-dependent cross sections within the VFNS have the same form as in the FFNS:
\begin{equation}\label{39}
\sigma^{(\mathrm{ACOT})}_{k}(x,\lambda)=\int\limits_{\chi}^{1}\text{d}z\,g(z,\mu_{F})\,
\hat{\sigma}_{k,\mathrm{g}}^{(0)}\!\left(x/z,\lambda\right) \quad  (k=L,A).
\end{equation} 

\begin{figure*}
\begin{center}
\begin{tabular}{ll}
\mbox{\epsfig{file=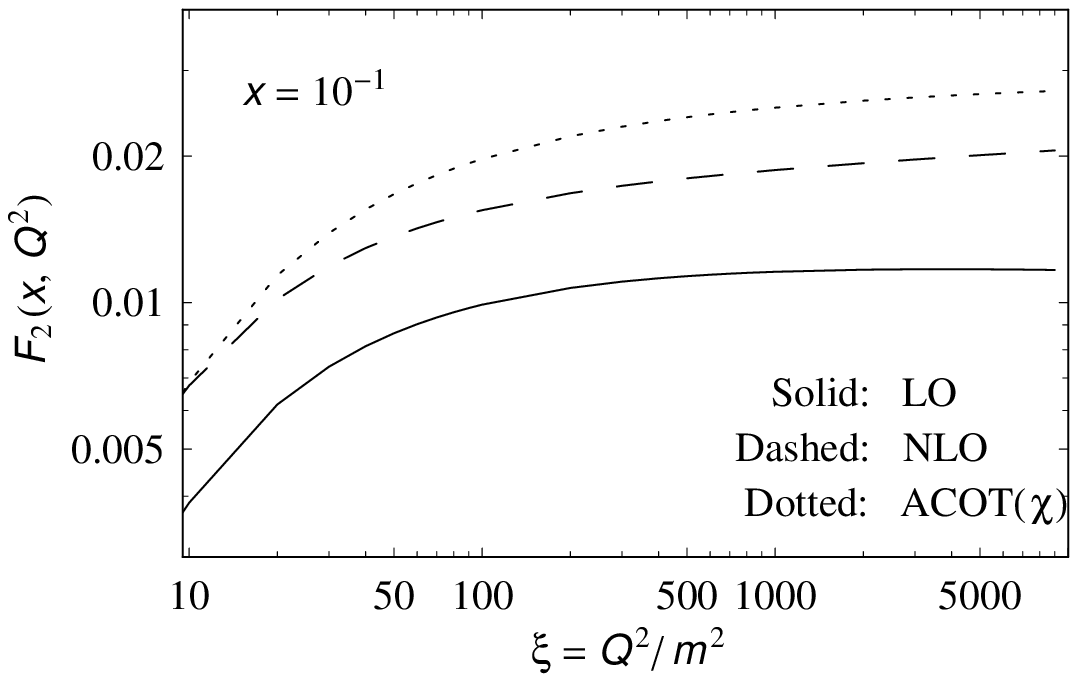,width=230pt}}
& \mbox{\epsfig{file=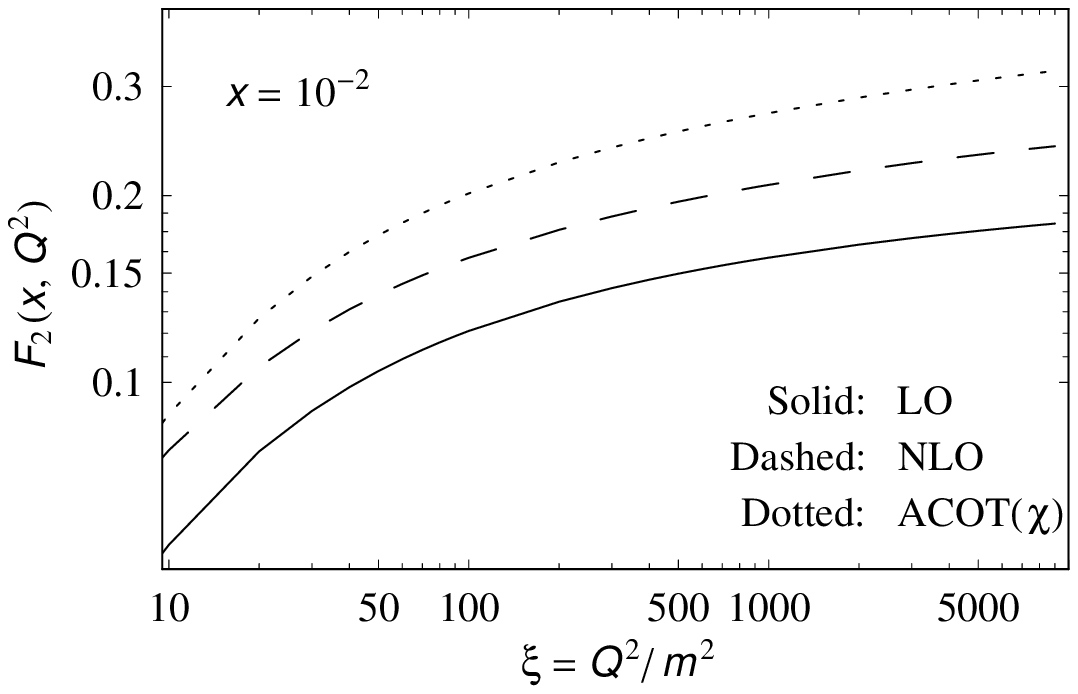,width=230pt}}
\end{tabular}
\caption{\label{Fg7}\small ${\cal O}(\alpha_{s})$ (solid lines), ${\cal O}(\alpha_{s}^2)$ (dashed lines) FFNS results and ${\cal O}(\alpha_{s})$ ACOT($\chi$) (dotted curves) predictions for $F_2(x,Q^2)$ in charm leptoproduction at $x=10^{-1}$ and $10^{-2}$.}
\end{center}
\end{figure*}

\begin{figure*}
\begin{center}
\begin{tabular}{ll}
\mbox{\epsfig{file=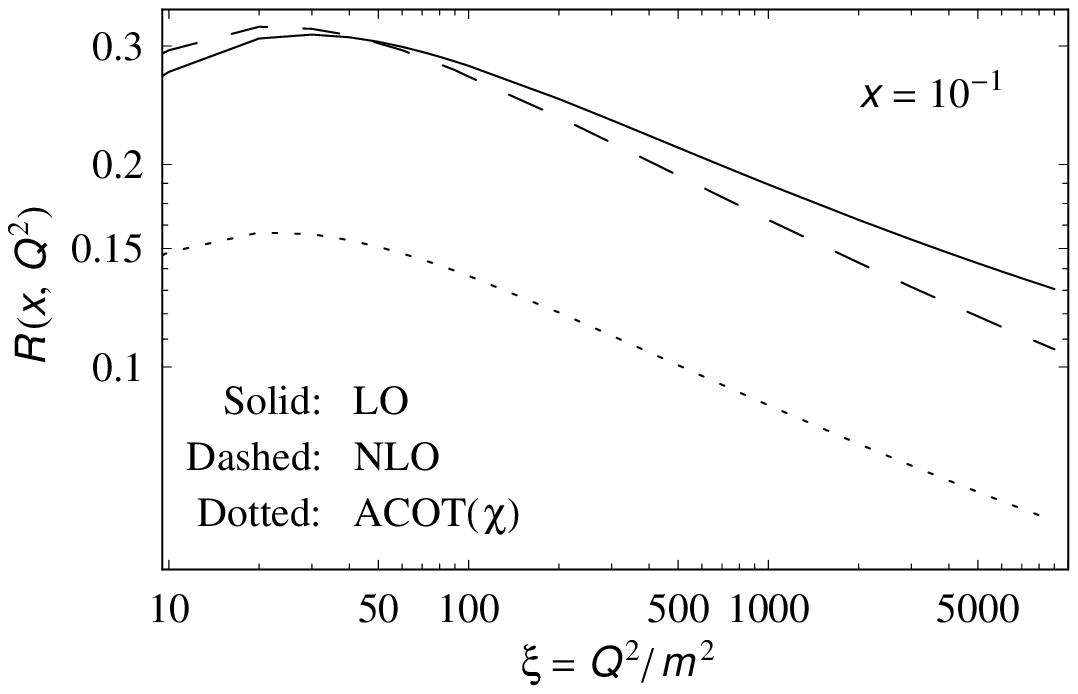,width=230pt}}
& \mbox{\epsfig{file=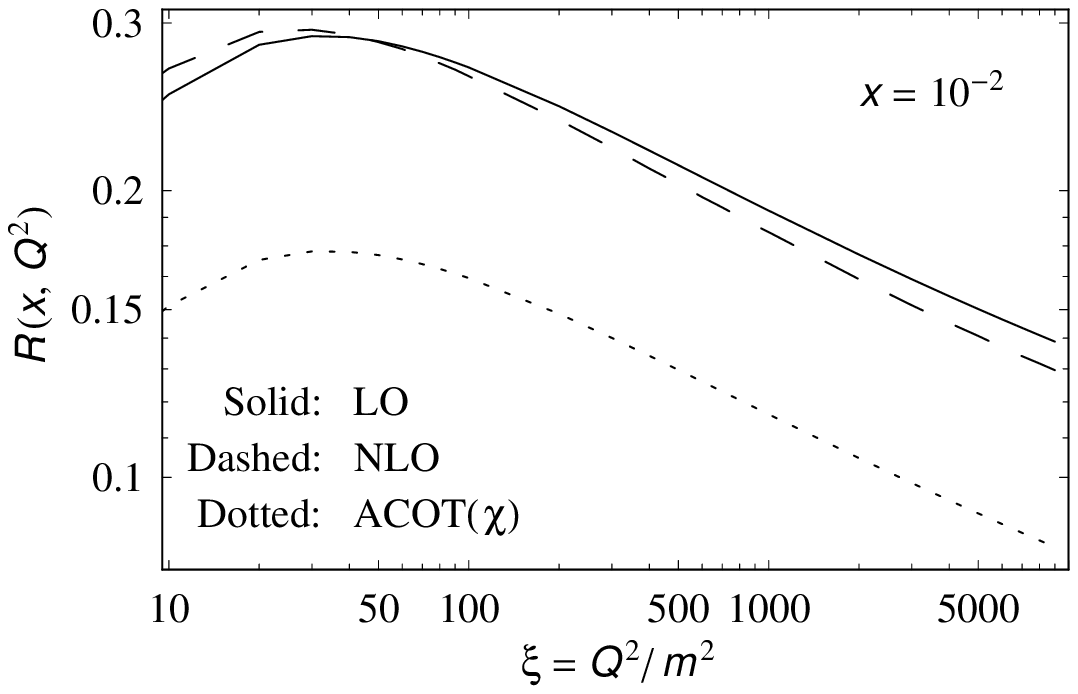,width=230pt}}
\end{tabular}
\caption{\label{Fg8}\small ${\cal O}(\alpha_{s})$ (solid lines), ${\cal O}(\alpha_{s}^2)$ (dashed lines) FFNS results and ${\cal O}(\alpha_{s})$ ACOT($\chi$) (dotted curves) predictions for $R(x,Q^2)$ in charm leptoproduction at $x=10^{-1}$ and $10^{-2}$.}
\end{center}
\end{figure*}

In Figs.~\ref{Fg7} and \ref{Fg8}, we present the ${\cal O}(\alpha_{s})$ and ${\cal O}(\alpha_{s}^2)$ FFNS predictions for the structure function $F_2(x,Q^2)$ and Callan-Gross ratio  $R(x,Q^2)=F_L/F_T$ in charm leptoproduction, and compare them with the corresponding ${\cal O}(\alpha_{s})$ ACOT($\chi$) results \cite{chi}. In our calculations, the CTEQ6M parameterization for PDFs and $m_c=1.3$~GeV for c-quark mass are used \cite{CTEQ6}. 

One can see from Fig.~\ref{Fg7} that both the radiative corrections and charm-initiated contributions to $F_{2}(x,Q^{2})$ are large: they increase the ${\cal O}(\alpha_{s})$ FFNS results by approximately a factor of two  at $x\sim 10^{-1}$ for all $Q^2$. At the same time, the relative difference between the dashed and dotted lines is not large: it does not exceed $25\%$ for $\xi=Q^2/m^2<10^{3}$. We conclude that it will be very difficult to determine the charm content of the proton using only data on $F_{2}(x,Q^{2})$ due to large radiative corrections (with corresponding theoretical uncertainties) to this quantity. 

Considering the corresponding predictions for the quantity $R(x,Q^2)$ presented in Fig.~\ref{Fg8}, we see that, in this case, the ${\cal O}(\alpha_{s}^2)$ FFNS and charm-initiated ${\cal O}(\alpha_{s})$ ACOT($\chi$) contributions are strongly different. In particular, the ${\cal O}(\alpha_{s}^2)$ FFNS corrections to $R(x,Q^2)$ are small, less than $15\%$, for $x\sim 10^{-2}$--$10^{-1}$ and $\xi<10^{4}$. 

At the same time, the ${\cal O}(\alpha_{s})$ charm-initiated contributions to $R(x,Q^2)$ are large: they decrease the ${\cal O}(\alpha_{s})$ FFNS predictions by about $50\%$ practically for all values of $\xi>10$. 
This is due to the fact that resummation of the mass logarithms has different effects on
the structure functions $F_{T}(x,Q^{2})$ and $F_{L}(x,Q^{2})$. In particular, contrary to 
the transverse structure function, the longitudinal one does not contain leading mass logarithms of the type $\alpha_s\ln (Q^{2}/m^{2})$ at both ${\cal O}(\alpha_{s})$ and ${\cal O}(\alpha_{s}^2)$ \cite{BMSMN}. For this reason, resummation of these logarithms within the VFNS leads to increasing of the quantity $F_{T}$ but does not affect the function $F_{L}$. We conclude that the Callan-Gross ratio $R(x,Q^2)=F_L/F_T$ could be good probe of the charm density in the proton at $x\sim 10^{-2}$--$10^{-1}$.
\begin{figure*}
\begin{center}
\begin{tabular}{ll}
\mbox{\epsfig{file=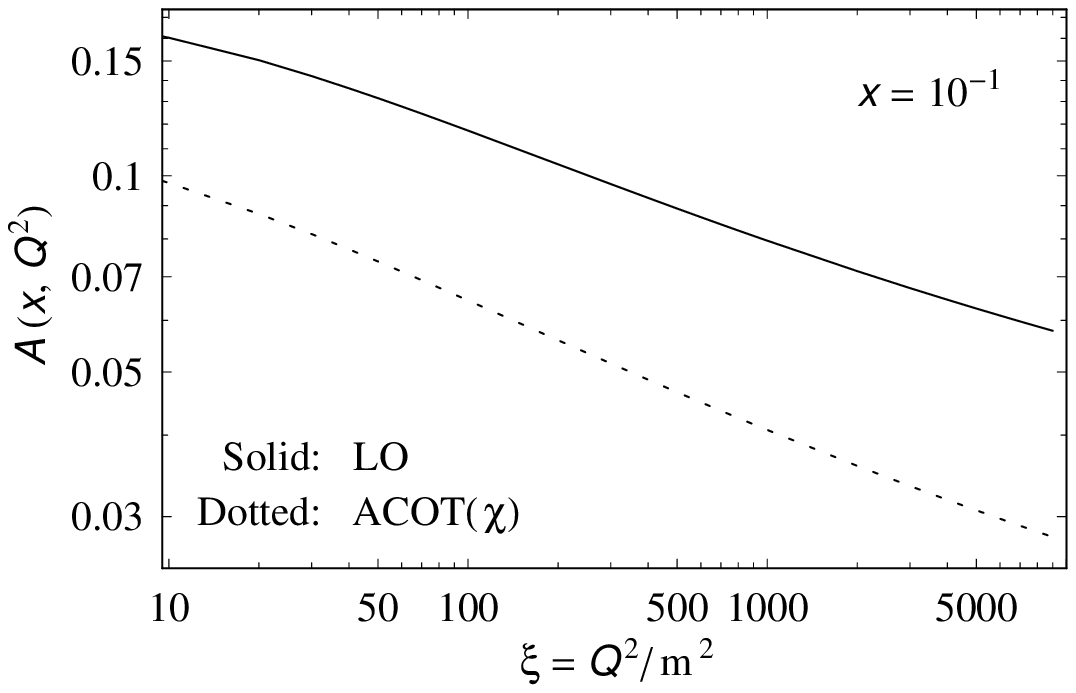,width=230pt}}
& \mbox{\epsfig{file=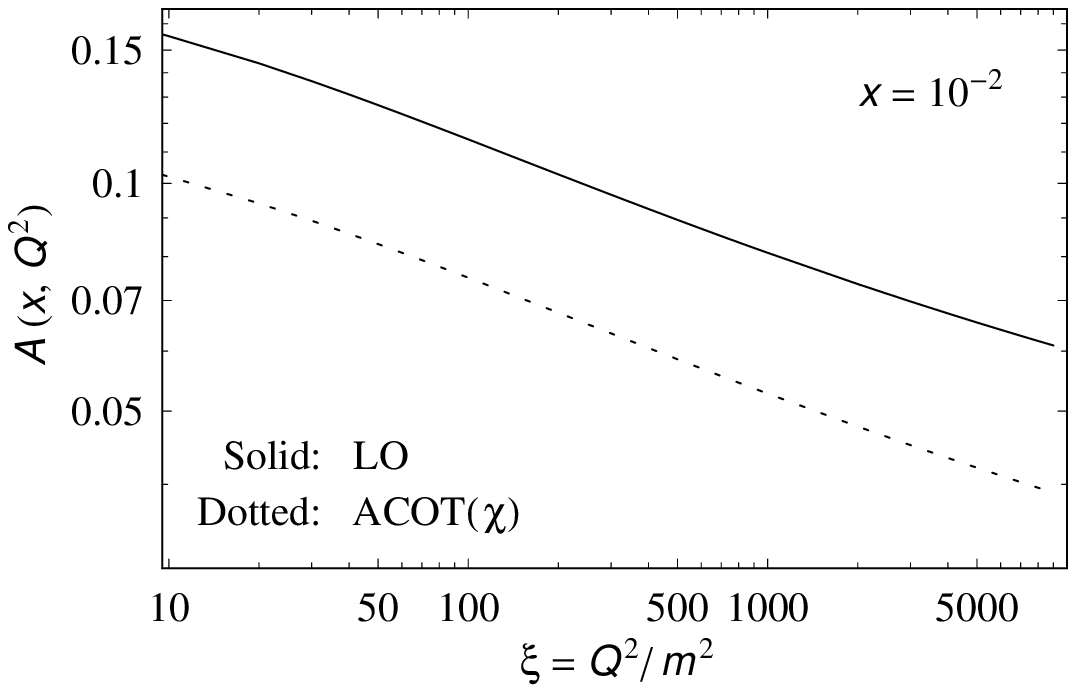,width=230pt}}
\end{tabular}
\caption{\label{Fg9}\small ${\cal O}(\alpha_{s})$ FFNS (solid lines) and ACOT($\chi$) (dotted curves) predictions for $A(x,Q^2)$ in charm leptoproduction at $x=10^{-1}$ and $10^{-2}$.}
\end{center}
\end{figure*}

Fig.~\ref{Fg9} shows the ${\cal O}(\alpha_{s})$ FFNS and ACOT($\chi$) predictions for the azimuthal asymmetry $A(x,Q^2)=2xF_{A}/F_{2}$ at $x=10^{-1}$ and $10^{-2}$.\footnote{We do not provide the radiative corrections for $A(x,Q^2)$ because the corresponding exact ${\cal O}(\alpha_{s}^2)$ predictions are not presently available while the soft-gluon approximation is unreliable for high $Q^{2}\gg m^{2}$.} One can see from Fig.~\ref{Fg9} that the mass logarithms resummation leads to a sizeable decreasing of the ${\cal O}(\alpha_{s})$ FFNS predictions for the $\cos2\varphi$ asymmetry. In the ACOT($\chi$) scheme, the charm-initiated contribution reduces the FFNS results for $A(x,Q^{2})$ by about $(30$--$40)\%$. The origin of this reduction is the same as in the case of $R(x,Q^{2})$: in contrast to $F_{2}(x,Q^{2})$, the azimuth-dependent structure function $F_{A}(x,Q^{2})$ is safe in the limit $m^2\to 0$. We see that the impact of the mass logarithms resummation on the $\cos2\varphi$ asymmetry is essential at $x\sim 10^{-2}$--$10^{-1}$ and therefore can be tested experimentally.
\begin{figure*}
\begin{center}
\begin{tabular}{ll}
\mbox{\epsfig{file=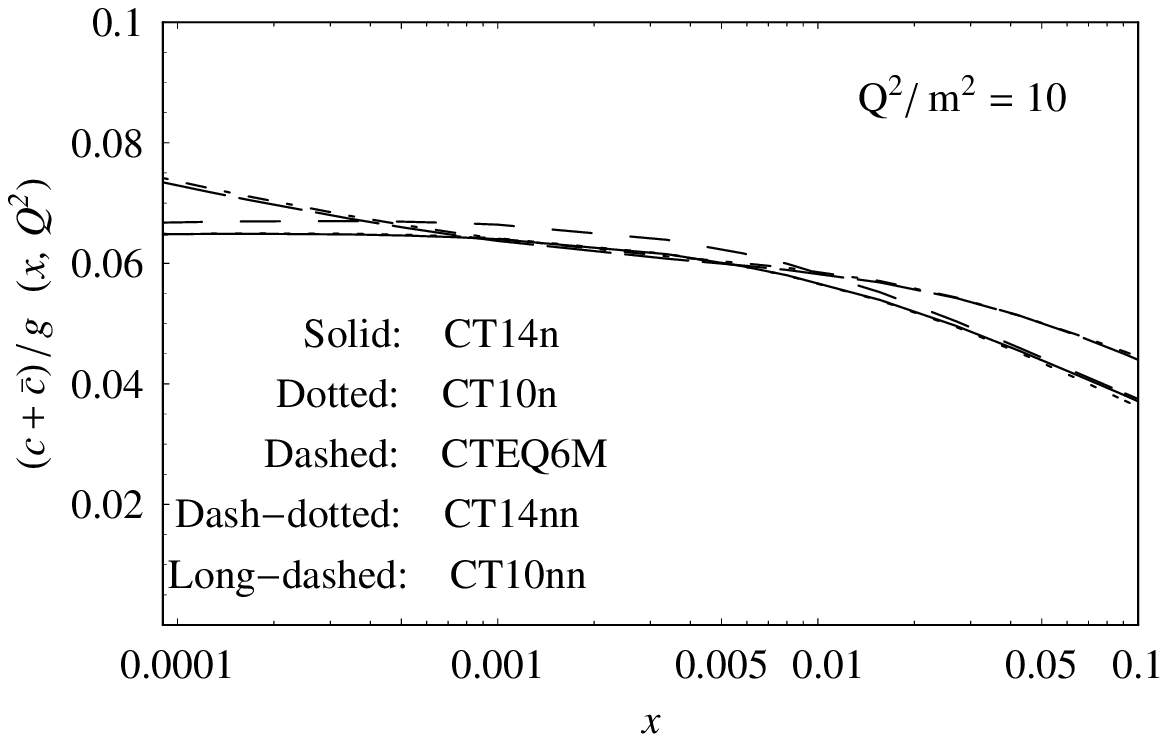,width=230pt}}
& \mbox{\epsfig{file=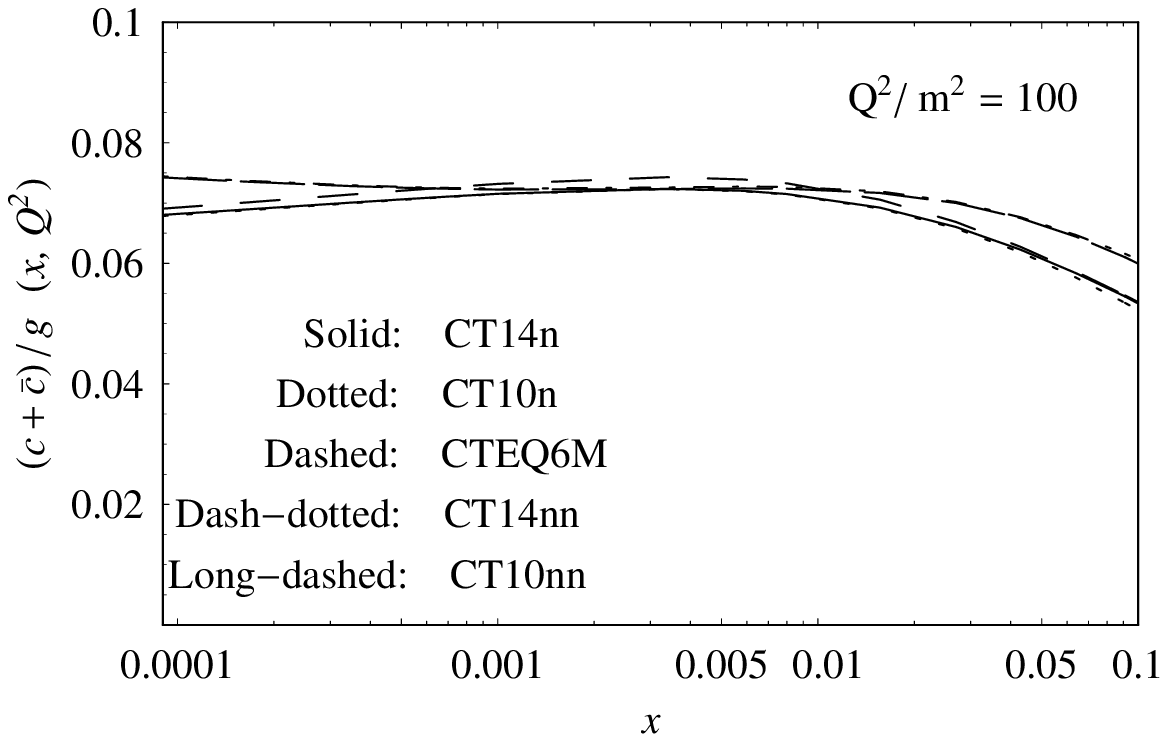,width=230pt}}
\end{tabular}
\caption{\label{Fg10}\small Predictions of the CT14n (solid line), CT10n (dotted line), CTEQ6M (dashed line), CT14nn (dash-dotted line) and CT10nn  (long-dashed line) versions of the PDFs for the quantity $(c+\bar{c})/g~(x,Q^2)$ at $Q^2/m^2=10$ and $10^2$.}
\end{center}
\end{figure*}

Note that our conclusions depend weakly on the PDFs in use since we analyze the ratios of the hadron-level cross sections. Moreover, one can see from Fig.~\ref{Fg10} that all the last "nlo" and "nnlo" sets of the CTEQ PDFs  \cite{CTEQ6,CT10,CT14} predict practically the same values for the charm content of the proton: $(c+\bar{c})/g~(x,Q^2)\approx$ (6--7)$\%$ in wide region of $x$ and $Q^{2}$.

In Figs.~\ref{Fg7}--\ref{Fg9}, we present the ${\cal O}(\alpha_{s})$ ACOT results for $R(x,Q^{2})$ and $A(x,Q^{2})$ as simplest illustrative examples. Our main conclusions about the resummation of mass logarithms are valid at both ${\cal O}(\alpha_{s})$ and ${\cal O}(\alpha_{s}^2)$. Indeed, a simple consideration of the FFNS LO and NLO results for the photon-gluon fusion \cite{LRSN,BMSMN} shows that the contribution of the leading mass logarithms to $F_2(x,Q^{2})$ has the form $\alpha_{s}^{n}\ln^{n}(Q^{2}/m^{2})$. The contribution of the leading mass logarithms to $F_L(x,Q^{2})$ is suppressed and has the form $(m^{2}/Q^{2})~\alpha_{s}^{n}\ln^{n}(Q^{2}/m^{2})$.\footnote{As to the subleading logarithms, $\alpha_{s}^{n}\ln^{n-1}(Q^{2}/m^{2})$, their resummation is expected to be suppressed by $\alpha_{s}$.}  Thus we can conclude that, contrary to $F_2(x,Q^{2})$, the resummation of mass logarithms for $F_L(x,Q^{2})$ is of subleading twist to all orders in $\alpha_{s}$. The same situation takes also place for $F_A(x,Q^{2})$.

We have verified this statement in first two orders of perturbation theory. One can see from Eqs.~(\ref{37}) that the lowest order QS subprocess is $\varphi$-independent, $\hat{\sigma}_{A,\mathrm{Q}}^{(0)}(z,\lambda)=0$, and has suppressed longitudinal component, $\hat{\sigma}_{L,\mathrm{Q}}^{(0)}(z,\lambda)\sim m^{2}/Q^{2}$.

In Ref.~\cite{we5}, the radiative corrections to the QS subprocess have been calculated. Our analysis shows that the ${\cal O}(\alpha_{s})$ predictions for both $\hat{\sigma}_{A,\mathrm{Q}}^{(1)}(z,\lambda)$ and $\hat{\sigma}_{L,\mathrm{Q}}^{(1)}(z,\lambda)$ are negligible for $Q^{2}/m^{2}\gg 1$.

So, we conclude that, contrary to the transverse component of the QS contribution, the 
longitudinal and azimuthal ones are of subleading twist to all orders in $\alpha_{s}$. This fact implies that resummation of the mass logarithms for the longitudinal and azimuth-dependent cross sections is, in principle, not necessary. For this reason, the VFNS predictions for $R(x,Q^{2})$ and $A(x,Q^{2})$ are smaller than the FFNS ones in both ${\cal O}(\alpha_{s})$ and ${\cal O}(\alpha_{s}^2)$. The difference between the FFNS and VFNS predictions for $R^c(x,Q^{2})$ is determined by the relative value of the charm density contribution to $F_2^c(x,Q^{2})$. The smaller the relative size of the charm-initiated contribution to $F_2^c(x,Q^{2})$, the less the difference between the FFNS and VFNS predictions for $R^c(x,Q^{2})$. The same situation takes also place for $A^c(x,Q^{2})$.

The ${\cal O}(\alpha_{s}^2)$ S-ACOT results presented in Refs.~\cite{sacot12-2,sacot12} clearly support our expectations. In particular, one can see from Fig.5 in Ref.~\cite{sacot12} that the difference between the VFNS and FFNS curves for $F_L^c(x,Q^{2})$ is of about 5$\%$ at $x=10^{-2}$ and $\xi\lesssim 10^2$.\footnote{At very high $Q^2$, there is ambiguity in separation of the heavy- and light-quark components for the structure functions, $F_{2,L}^{l,c}(x,Q^{2})$ within the ${\cal O}(\alpha_{s}^2)$ VFNS. For this reason, the definition (42) in Ref.~\cite{sacot12}  for heavy-quark components may be inappropriate at $\xi > 10^2$.} The corresponding difference for $F_2^c(x,Q^{2})$ at ${\cal O}(\alpha_{s}^2)$ is of about (15--20)$\%$. Using $F_L^{\mathrm{VFNS}}(x,Q^{2})=F_L^{\mathrm{FFNS}}(x,Q^{2})$, one can obtain from ~\cite{sacot12}   that resummation of the mass logarithms reduces the FFNS results for $R^c$ within the ${\cal O}(\alpha_{s}^2)$ S-ACOT scheme by about 20$\%$. Remember, that the corresponding reduction within the ${\cal O}(\alpha_{s})$ ACOT($\chi$) approach is of about 50$\%$. 

We see that the ${\cal O}(\alpha_{s}^2)$ FFNS and VFNS predictions for $R^c(x,Q^{2})$ are closer to each other than the ${\cal O}(\alpha_{s})$ ones. This fact is in accordance with our expectations because the relative contribution of the charm-initiated component to $F_2^c(x,Q^{2})$ at ${\cal O}(\alpha_{s}^2)$ is less than at ${\cal O}(\alpha_{s})$. Indeed, while the ratio $(c+\bar{c})/g~(x,Q^2)$ is practically the same in both "nlo" and "nnlo" sets of available PDFs, the ${\cal O}(\alpha_{s}^2)$ predictions for $F_2^c(x,Q^{2})$ contain sizable light-quark initiated contributions which are absent at ${\cal O}(\alpha_{s})$.  

\section{Conclusion}

We conclude by summarizing our main observations. In the present paper, we first review the available theoretical results for the Callan-Gross ratio, $R(x,Q^{2})$, and azimuthal $\cos(2\varphi)$ asymmetry, $A(x,Q^{2})$, in heavy-quark leptoproduction. 
It turned out that large (especially, at non-small $x$) radiative corrections to the structure functions cancel each other in their ratios $R(x,Q^2)=F_L/F_T$ and $A(x,Q^{2})=2xF_A/F_2$ with good accuracy. As a result, the ${\cal O}(\alpha_{s}^2)$ contributions to the ratios $R(x,Q^{2})$ and $A(x,Q^{2})$ do not exceed $10$--$15\%$ in a wide region of the variables $x$ and $Q^2$. Our analysis shows that, sufficiently above the production threshold, the pQCD predictions for $R(x,Q^2)$ and $A(x,Q^{2})$ are insensitive (to within ten percent) to standard uncertainties in the QCD input parameters and to the DGLAP evolution of PDFs. We conclude that, unlike the production cross sections, the Callan-Gross ratio and $\cos(2\varphi)$ asymmetry in heavy-quark leptoproduction are quantitatively well defined in pQCD. Measurements of the quantities $R(x,Q^2)$ and $A(x,Q^{2})$ in charm and bottom leptoproduction would provide a good test of the conventional parton model based on pQCD.

Then we discuss some experimental and phenomenological applications of the observed perturbative stability. Our main conclusion is that the quantities $R(x,Q^{2})$ and $A(x,Q^{2})$ will be good probes of the heavy-quark densities in the proton. 

The VFN schemes have been proposed to resum the mass logarithms of the form $\alpha_{s}^{n}\ln^{n}(Q^{2}/m^{2})$ which dominate the production cross sections at high energies, $Q^2\to \infty$. Evidently, were the calculation done to all orders in $\alpha_{s}$, the VFNS and FFNS would be exactly equivalent. There is a point of view advocated in Refs.~\cite{ACOT,Collins} that, at high energies, the perturbative series converges better within the VFNS than in the FFNS. There is also another opinion \cite{Stratmann,BMSN,Neerven} that the above logarithms do not vitiate the convergence of the perturbation expansion so that a resummation is, in principle, not necessary. Our analysis indicates two promising  experimental ways to resolve this problem: using the Callan-Gross ratio and/or azimuthal $\cos(2\varphi)$ asymmetry in DIS. The quantities $R(x,Q^2)$ and $A(x,Q^2)$ are perturbatively stable in the FFNS  but sensitive to resummation of the mass logarithms of the type $\alpha_{s}\ln\left( Q^{2}/m^{2}\right)$ within the VFNS. Our analysis shows that resummation of the mass logarithms leads to reduction of the ${\cal O}(\alpha_{s})$ FFNS predictions for $A(x,Q^2)$ and $R(x,Q^2)$ by $(30$--$50)\%$ at $x\sim 10^{-2}$--$10^{-1}$ and $Q^2\gg m^2$.\footnote{Within the ${\cal O}(\alpha_{s}^2)$ S-ACOT VFNS \cite{sacot12}, the corresponding reduction of the FFNS predictions for $R^c(x,Q^2)$ is estimated to be (15--20)$\%$.} Therefore measurements of the ratios $R(x,Q^2)$ and $A(x,Q^2)$ in heavy-qaurk leptoproduction would make it possible to clarify the question whether the VFNS perturbative series is more reliable than the FFNS one. 

As to the experimental aspects, the Callan-Gross ratio and azimuthal $\cos(2\varphi)$  asymmetry in heavy-flavor leptoproduction can be measured in the current COMPASS and proposed EIC \cite{EIC} and LHeC \cite{LHeC} experiments.

\begin{acknowledgements}
The author is grateful to S.~J.~Brodsky, A.~V.~Efremov, A.~V.~Kotikov, A.~B.~Kniehl, E.~Leader, S.~O.~Moch, A.~G.~Oganesian O.~V.~Teryaev and C.~Weiss for useful discussions. We thank S.~I.~Alekhin and J.~Bl\"umlein for providing us with fast code \cite{Blumlein} for numerical calculations of the NLO partonic cross sections. This work is supported in part by the State Committee of Science of RA, grant 15T-1C223.
\end{acknowledgements}

\end{document}